\documentclass[fleqn]{article}
\pdfoutput=1
\usepackage{a4wide,epsfig}
\usepackage{epsfig,xspace}
\usepackage{rotating,amsmath,amssymb}
 \usepackage[greek,english]{babel}

\def\nue{\ensuremath{\nu_{e}}\xspace}
\def\nubare{\ensuremath{\overline{\nu}_{e}}}

\def\numu{\ensuremath{\nu_{\mu}\xspace}}
\def\nubarmu{\ensuremath{\overline{\nu}_{\mu}}}

\def\nutau{\ensuremath{\nu_{\tau}\xspace}}

\newcommand{\nuenumu}{\ensuremath{\nue \rightarrow \numu\xspace}}
\newcommand{\numunutau}{\ensuremath{\numu \rightarrow \nutau\xspace}}
\newcommand{\nuenutau}{\ensuremath{\nue \rightarrow \nutau}}

\newcommand{\nubarmunubare}{\ensuremath{\overline{\nu}_\mu \rightarrow \overline{\nu}_e\xspace}}
\newcommand{\dmot}{\ensuremath{\Delta m^2_{12}\xspace}}
\newcommand{\dmtt}{\ensuremath{\Delta m^2_{23} \xspace}}

\newcommand{\He}{\ensuremath{^6{\mathrm{He}\xspace}}}
\newcommand{\Ne}{\ensuremath{^{18}{\mathrm{Ne}\xspace}}}
\def\Li{^6{\mathrm{Li}}}
\def\Li8{\ensuremath{^8{\mathrm{Li}}}}
\def\B8{\ensuremath{^8{\mathrm{B}}}}

\def\anumu{\overline{{\mathrm\nu}}_{\mathrm \mu}}
\newcommand{\thetaot}{\ensuremath{\theta_{13}}\xspace}
\newcommand{\thetatt}{\ensuremath{\theta_{23}}\xspace}
\newcommand{\numunue}{\ensuremath{\nu_\mu \rightarrow \nu_e}\xspace}

\newcommand{\pnumunue}{\ensuremath{P(\nu_\mu \rightarrow \nu_e)}\xspace}

\newcommand{\sigdm}{\ensuremath{{\rm sign}(\Delta m^2_{23})\xspace}}
\newcommand{\delCP}{\ensuremath{\delta_{\rm CP}\xspace}}

\begin{document}

\thispagestyle{empty}
\vspace*{1cm}
\begin{center}
{\Large{\bf European facilities for accelerator neutrino physics: perspectives for the decade to come}
}\\
\vspace{.5cm}
R.~Battiston$^{\rm a}$, M.~Mezzetto$^{\rm b}$, P. Migliozzi$^{\rm c}$,  F.~Terranova$^{\rm d}$ \\
\vspace*{0.5cm}
$^{\rm a}$ Dep. of Physics, Univ. of Perugia and I.N.F.N., Sezione di Perugia, Perugia, Italy \\
$^{\rm b}$ I.N.F.N., Sezione di Padova, Padova, Italy  \\
$^{\rm c}$ I.N.F.N., Sezione di Napoli, Naples, Italy  \\
$^{\rm d}$ I.N.F.N., Laboratori Nazionali di Frascati,
Frascati (Rome), Italy \\
\end{center}

\begin{abstract}
Very soon a new generation of reactor and accelerator neutrino
oscillation experiments - Double Chooz, Daya Bay, Reno and T2K - will
seek for oscillation signals generated by the mixing parameter
\thetaot.  The knowledge of this angle is a fundamental milestone to
optimize further experiments aimed at detecting CP violation in the
neutrino sector. Leptonic CP violation is a key phenomenon that has
profound implications in particle physics and cosmology but it is
clearly out of reach for the aforementioned experiments.  Since late
90's, a world-wide activity is in progress to design facilities that
can access CP violation in neutrino oscillation and perform high
precision measurements of the lepton counterpart of the
Cabibbo-Kobayashi-Maskawa matrix. In this paper the status of these
studies will be summarized, focusing on the options that are best
suited to exploit existing European facilities (firstly CERN and the
INFN Gran Sasso Laboratories) or technologies where Europe has a world
leadership.  Similar considerations will be developed in more exotic
scenarios - beyond the standard framework of flavor oscillation among
three active neutrinos - that might appear plausible in the occurrence
of anomalous results from post-MiniBooNE experiments or the CNGS.
\noindent
\end{abstract}

\section{Open questions in neutrino oscillations and a glimpse to future}
\label{sec:intro}

The results of atmospheric, solar, accelerator and reactor neutrino
experiments \cite{giunti} accumulated in the last decades show that
flavor mixing occurs not only in the hadronic sector, as it has been
known for long, but in the leptonic sector as well. This outstanding
result has been obtained exploiting a quantum interference phenomenon
speculated by Pontecorvo in 1957 and named - in analogy with
the oscillation of neutral K mesons - ``neutrino
oscillation''~\cite{pontecorvo}. In its modern formulation, neutrino
oscillations occur because flavor states generated by weak
interactions, i.e. what we call ``electron neutrinos'' (\nue), ``muon
neutrinos'' (\numu) and ``tau neutrinos'' (\nutau), are not stationary
states but, indeed, a coherent superposition of states with rest mass $m_1$, $m_2$
and $m_3$. As a consequence, flavor evolves with time and neutrinos of
a given flavor $\nu_\alpha$ produced from a source at a distance $L$
from the detector, might be measured to have a different flavor
$\nu_\beta$. Pontecorvo noted in 1967 that, assuming only two flavors
(at that time the tau lepton and the \nutau\ were still unknown), the probability
of observing a \nue\ from a pure source of \numu\ is given by:
\begin{equation}
\pnumunue = \sin^2 2\theta  \sin^2 1.27 \frac{ \Delta m^2 ([eV^2]) L([km])}
{E([GeV])} \ ,
\label{eq:pontecorvo}
\end{equation}
being $\theta$ the angle that parametrizes a rotation between the
stationary state $\nu_1$ and $\nu_2$ and the flavor states \nue\ and
\numu , while  $\Delta m^2 \equiv m_2^2-m_1^2$ represents the
squared mass difference between $m_1$ and $m_2$. Neutrino oscillations
occur if neutrinos are massive particles, if their masses are non
degenerate ($m_1 \neq m_2$) and if $\theta\neq 0$. Nature has
fulfilled for us all these conditions and neutrino oscillations are
actually observables in many systems: in neutrinos produced from the sun, in
neutrinos generated by cosmic ray interactions with the earth
atmosphere and, more recently, in artificial neutrinos produced by
nuclear reactors and accelerators.

All experimental results gathered so far are mutually consistent if we
assume just three active neutrinos - the ones observed in a direct
manner in 1956 (\nue \cite{reines}), 1962 (\numu \cite{steinberger}),
and 2001 (\nutau \cite{donut}) - and if we describe such flavor states
as superposition of three mass eigenstates labeled $m_1$, $m_2$ and
$m_3$, respectively~\cite{bettini}.  As for the case of quarks, mass
eigenstates and flavor eigenstates do not coincide and the unitary
matrix that transform the former to the latter is non-diagonal. In the
quark case, this matrix is called the Cabibbo-Kobayashi-Maskawa
(CKM~\cite{CKM}) matrix and can be parametrized by three Euler angles
$\theta_{12},\theta_{23}$ and $\theta_{13}$ and one ``complex phase''
$\delta$\footnote{The parameter $\delta$ is actually a real number
that describes the phase factor $e^{i\delta}$ appearing in the mixing
matrix.}, the first angle $\theta_{12}$ being the ``Cabibbo angle''
$\theta_C$. In the quark sector we know for sure that $\delta$ is
different from zero: it causes CP violation in hadrons as observed
e.g. in the neutral kaon decays since 1964~\cite{cronin} or in the
neutral B decays since 2001~\cite{babar-belle}.  For leptons, the
corresponding matrix is sometimes referred as the
Pontecorvo-Maki-Nakagawa-Sakata ($U_{PMNS}$) mixing
matrix~\cite{pontecorvo,neutrino_osc}. If neutrinos are Dirac particle
as the quarks - i.e. neutrinos and antineutrinos are different
particles - $U_{PMNS}$ is, again, a $3\times 3$ unitary matrix and its
degrees of freedom are the same as for the CKM (three Euler angles
bounded between 0 and $\pi/2$ and one complex CP violating phase
$\delta$ ranging from 0 to $2\pi$). Two additional complex phases
appear if neutrinos are Majorana particles, i.e.  if neutrinos and
antineutrinos are the same particle; these ``Majorana phases'' are,
however, unobservable in oscillations~\cite{giunti} and will not be
considered hereafter.  Moreover, $U_{PMNS}$ is not necessarily
unitary: in future - and in a way similar to the CKM - direct
unitarity tests will likely become an active subject of empirical
investigation.

Neutrino oscillations are a powerful tool to determine the squared
mass differences of the rest masses of neutrinos and the four
parameters $\theta_{12},\theta_{23}$, $\theta_{13}$ and $\delta$.  The
experimental results obtained so far point to two very distinct mass
differences, $\Delta m^2_{sol} = \dmot \equiv m^2_2 -m^2_1 =
7.65^{+0.23}_{-0.20} \times 10^{-5}$ eV$^2$ and $|\Delta m^2_{atm}| =
|\dmtt| \equiv |m^2_3 -m^2_2| \simeq |m^2_3 -m^2_1| =
2.40^{+0.12}_{-0.11} \times 10^{-3}$ eV$^2$
\cite{SchwetzTortolaValle}.  In jargon, \dmot\ is called the ``solar
mass scale'' because it drives oscillation of solar neutrinos, but, of
course, if the energy of the neutrino and the source-to-detector
distance are properly tuned, it can be observed also using man-made
neutrinos, e.g. reactor neutrinos located about 100~km from the
detector (as it was the case for KAMLAND~\cite{kamland}). Similarly,
the standard framework predicts that atmospheric neutrinos mainly
oscillate at a frequency that depends on \dmtt (``atmospheric
scale''). Accelerator neutrinos can see (actually, saw in
K2K~\cite{k2k} and MINOS~\cite{minos}) the same effect using neutrinos
of energy around 1~GeV and source-to-detector distances
(``baselines'') of a few hundreds of km (250~km in K2K and 730~km in
MINOS).  Only two out of the four parameters of the leptonic mixing
matrix are known: $\theta_{12} = 33.5 ^{+1.3}_{-1.0}$ degrees and
$\theta_{23} = 42^{+4}_{-3}$ degrees \cite{SchwetzTortolaValle}.
Surprisingly, these values are much larger than for the CKM
($\theta_{12} \approx 13^\circ$ and $\theta_{23}\approx
2.3^\circ$). The other two parameters, $\theta_{13}$ and $\delta$, are
still unknown: for the mixing angle $\theta_{13}$ direct searches at
reactors (the CHOOZ~\cite{chooz} and Palo Verde~\cite{palo_verde}
experiments) combined with solar neutrino data give the upper bound
$\theta_{13} \leq 11.5^\circ$ at 90\% CL, whereas for the leptonic
CP-violating phase $\delta$ we have no information whatsoever.  In the
standard framework, atmospheric neutrinos mainly oscillate from
$\numu$ to $\nutau$ and vice-versa with the \dmtt-driven frequency.
All other transitions are sub-dominant and depend on the parameter
\thetaot. The probability of \numunue transitions at the atmospheric
scale is, at leading order, $\simeq 0.5 \sin^2 2\theta_{13} \sin^2
\left[ 1.27 \dmtt L/E \right]$, so its smallness is a consequence of
the smallness of \thetaot in $U_{PMNS}$.  All these considerations
hold in the standard framework, i.e. under the assumption that
atmospheric neutrinos disappear due to \numunutau\ oscillation.  This
is a well-grounded assumption, which is however being tested in a very
straightforward manner (observation of \nutau\ charged-current
interactions) at the CERN-to-Gran Sasso (CNGS) neutrino
beam~\cite{opera_first}.

Reactor antineutrinos are a classical tool to probe the size of
\thetaot since they have an energy of a few MeV and the ratio $L/E$
matches the atmospheric scale, i.e. $1.27 \dmtt L/E \simeq
\pi/2$, at baselines of about 1 km. If they oscillate toward tau or
muon antineutrinos, an apparent flux reduction in the detector will be
observed.  Currently, the best direct experimental limit on
\thetaot comes from the CHOOZ reactor experiment. A world
limit can be derived~\cite{Schwetz:2008er} by a full $3\nu$ analysis
of all the neutrino oscillation experiments, see Tab.~\ref{tab:th13}.
Such world limit provides a slightly looser value than the
CHOOZ limit. It is an indication that the best fit for \thetaot might
be different from zero \cite{Fit th13}, although with small
statistical significance ($< 2\sigma$).

\begin{table}[h]
{$
  \sin^2\theta_{13} \le \left\lbrace \begin{array}{l@{\qquad}l}
      0.049~(0.072) & \rm{(solar+KamLAND)} \\
      0.033~(0.051) & \rm{(Chooz+atm+K2K+MINOS)} \\
      0.032~(0.046) & \rm{(global\; 2008 data)} \\
      0.04~(0.05) & \rm{(global\; 2009 data)} \\
    \end{array} \right.
$}
\caption{The 90\% CL ($3\sigma$) bounds on $\sin^2\theta_{13}$ from
an analysis of different sets of data  \protect\cite{Schwetz:2008er} }
\label{tab:th13}
\end{table}

A preliminary analysis of the MINOS experiment \cite{Minos-nue} shows
a $1.5 \sigma$ excess of \nue-like events in the far detector, that
could be interpreted as a manifestation of non-zero value of
\thetaot.  Very recently, however, the Super-Kamiokande
Collaboration~\cite{SK-Taup09} presented an updated analysis of its
full atmospheric neutrino sample: their data do not support the above hint for
\thetaot $\ne 0$.

A non-zero value of the mixing angle \thetaot plays an essential role
in neutrino oscillation physics.  Since late 90's, it has been
realized that the oscillation \numunue at the atmospheric scale has a
very rich structure. Its leading term depends on $\sin^2 2\thetaot$
but other subdominant terms suppressed by power of the ratio
\dmot/\dmtt, contain explicit dependence on the CP phase $\delta$.
Moreover, neutrino propagation from the source to the detector through
the earth is affected by matter effects similar to the ones perturbing
the free streaming of the solar neutrinos~\cite{msw}. It is a
remarkable fact that such perturbations are sensitive not only to the
overall size of \dmtt\ but also to its sign, i.e. to the relative
ordering of $m_1, m_2$ and $m_3$.  In the most used parametrization of
the leptonic mixing~\cite{PDG} $m_2 > m_1$ by definition and $m_3$ can
be either larger (``normal hierarchy'') or smaller
(``inverted hierarchy'') than $m_2$, giving $\sigdm >0$ and $\sigdm
<0$ respectively. Again, this information might be retrieved from high
precision experiments measuring both \numunue\ and \nubarmunubare\
transitions.  The unique role of \numunue \, and its CP conjugate
\nubarmunubare \, at the atmospheric scale is a consequence of the
fact that the ratio \dmot/\dmtt \, is not too small (about 1/30) and
three-family interference effects are sizable. Hence, subdominant
three-family interference amplitudes are observables with neutrinos in
analogy to the $K^0-\bar{K}^0$ system for quarks. Again, this
occurrence is fortunate and purely accidental. Before 2001, several
other values, much smaller than $\dmot \simeq 7.9 \times 10^{-4}$ eV$^2$ were
still experimentally allowed. The demonstration that \dmot\ is in the
ballpark of $10^{-3}$~eV$^2$ is considered the real inception of the
precision era of neutrino physics, since it opened up the possibility
to measure the whole of $U_{PMNS}$ with artificial sources, save, once
more, a finite value for \thetaot.

To give a feeling of the complexity of the task of studying
subdominant \numunue transitions we report below the approximate
formula that holds when matter effects are not negligible but matter density
is constant along the neutrino path. In this case the
transition probability $\nu_e \to \nu_\mu$ ($\bar \nu_e \to \bar
\nu_\mu$) can
be written as~\cite{PilarNufact}:
\begin{equation}
  P^\pm (\nu_e \to \nu_\mu)=
   X_\pm \sin^2 (2 \theta_{13}) + Y_\pm \cos ( \theta_{13} ) \sin (2 \theta_{13} )
  \cos \left ( \pm \delta - \frac{\Delta m^2_{23} L }{4 E_\nu} \right ) + Z \, ,
\label{eq:Donini}
\end{equation}
where $\pm$ refers to neutrinos and antineutrinos, respectively and
$a[{\rm eV}^2]=\pm 2\sqrt{2}G_Fn_eE_\nu=7.6 \times 10^{-5}\rho[g/cm^3]E_\nu[{\rm GeV}]$
is the electron density in the material crossed by neutrinos.
The coefficients of the two equations are:
$
X_\pm = \sin^2 (\theta_{23} )
\left ( \frac{\Delta m^2_{23} }{ | a - \Delta m^2_{23}| } \right )^2 \sin^2 \left ( \frac{|a - \Delta m^2_{23}| L}{ 4 E_\nu } \right )
$, \\
$
Y_\pm = \sin ( 2 \theta_{12} ) \sin ( 2 \theta_{23} )
\left ( \frac{\Delta m^2_{12} }{ a } \right ) \left ( \frac{\Delta m^2_{23} }{ |a - \Delta m^2_{23}| } \right )
\sin \left ( \frac{ a L }{ 4 E_\nu } \right ) \sin \left ( \frac{ |a - \Delta m^2_{23}| L }{ 4 E_\nu } \right )
$,\\
$
Z = \cos^2 (\theta_{23} ) \sin^2 (2 \theta_{12})
\left ( \frac{\Delta m^2_{12} }{ a } \right )^2 \sin^2 \left ( \frac{a L }{ 4 E_\nu } \right )
$.

The \numunue\ transitions are driven by the \thetaot term which is
proportional to $\sin^2{2\thetaot}$ and $P(\nu_\mu \rightarrow \nu_e)$
could be strongly influenced by the unknown value of \delCP \, and
\sigdm.

Given the complexity of the \numunue\  transition formula  it will be very
difficult for pioneering experiments to extract all
the unknown parameters unambiguously. Correlations are present between \thetaot and
$\delta$ \cite{PilarNufact}. Moreover, in absence of information about
 the sign of $\Delta m^2_{23}$~\cite{MinakataDege,Barger:2001yr} and the
approximate $[\theta_{23}, \pi/2 - \theta_{23}]$ symmetry for the
atmospheric angle~\cite{FogliDege}, additional clone solutions rise up. In general,
the measurement of $P(\nu_\mu \to \nu_e)$ and $P(\bar \nu_\mu \to \bar
\nu_e)$ will result in eight allowed regions of the parameter space,
the so-called eightfold-degeneracy~\cite{Barger:2001yr}. It can be solved
only combining several measurements at different energies or baseline, or exploiting channels
different from  \numunue\ , as \nuenutau.

\begin{figure}
\centering
{\includegraphics[width=12cm]{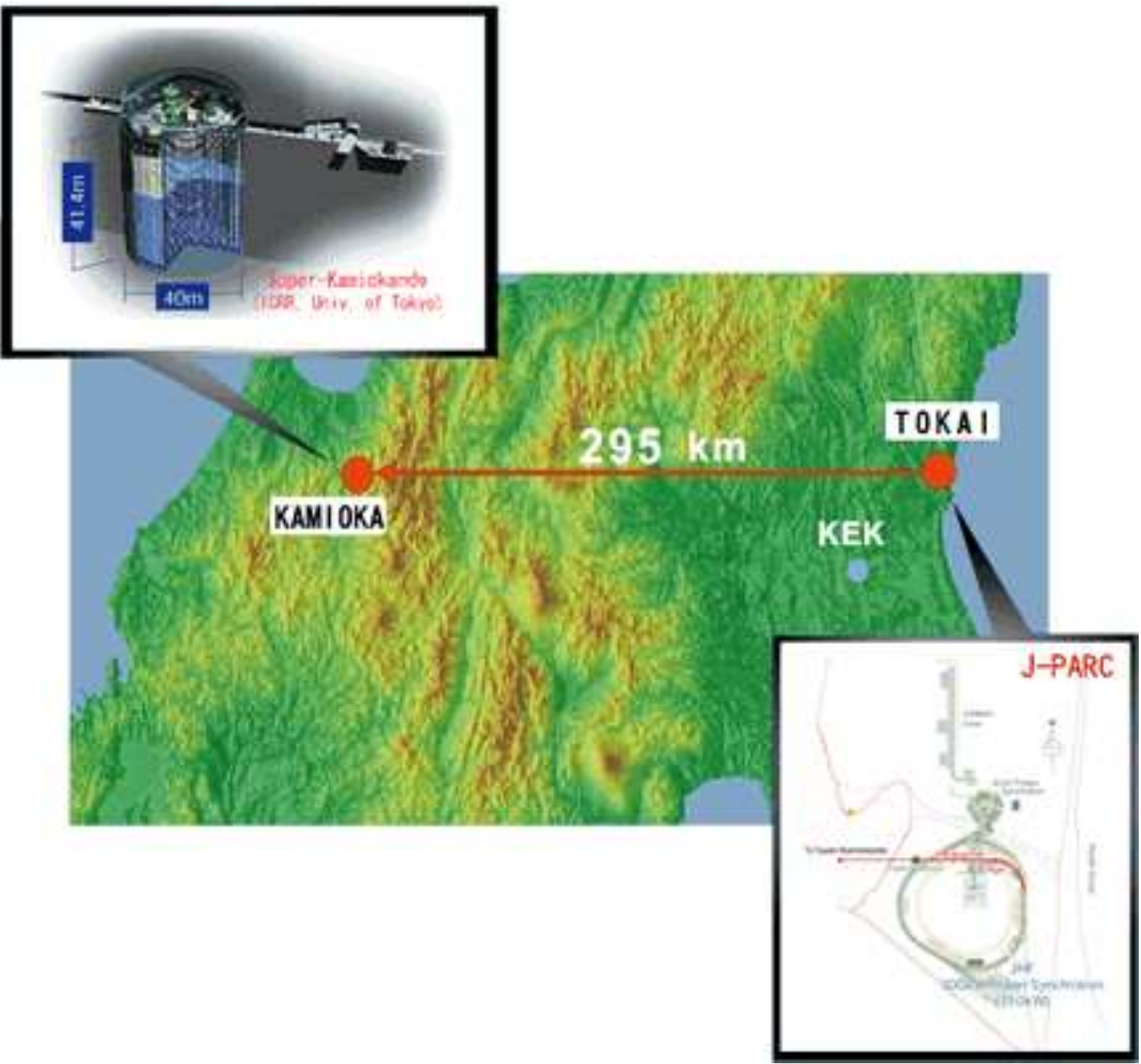}
    \caption{The T2K experiment exploits a new neutrino beam of 0.75~MW
maximum power built into the JPARC acceleration complex, at the Tokai
campus of the Japan Atomic Energy Agency. Neutrinos of mean energy of
0.7~GeV are sent toward the 13-year-old SuperKamiokande detector (22000 ton fiducial
mass), located 292~km far from JPARC.}  }
    \label{fig:T2K}
\end{figure}

This is the reason why it is commonly believed that the next
generation of experiments will mainly focus on the very evidence of
\nue appearance at the atmospheric scale, while the task of precision
measurement, including CP violation and determination of the mass
hierarchy, will take a further round of
facilities~\cite{Huber:2009cw,workshopCERN}. At present, there is only one
accelerator experiment in data taking - actually in the startup phase
- that has been optimized for the search of non-null \thetaot.  It is
the T2K~\cite{T2K} experiment in Japan, which is expected to improve
by an order of magnitude the limit from CHOOZ exploiting a new beam
from JPARC to the pre-existing far detector of SuperKamiokande
(Fig.~\ref{fig:T2K}).  The relevance of \thetaot search to ground the
future of the field is such that it boosted a large number of
proposals to improve CHOOZ with novel detectors at reactors. Among the
$\sim10$ original proposals, three of them survived and are either in
commissioning or in construction phase. The most advanced is
Double-Chooz~\cite{DoubleChooz}, in the original CHOOZ site in France,
which makes use of a larger far detector and of a new near detector
identical to the far detector for flux normalization (a near detector
was not available at the time of CHOOZ).  Double-Chooz will start data
taking in a few months with the far detector only; the near detector
will be available after 2011.  The other reactor experiments under
construction are Daya-Bay~\cite{DayaBay} in China and RENO~\cite{RENO}
in Korea. In particular, the basic experimental layout of Daya Bay
consists of three underground experimental halls, one far and two
near, linked by horizontal tunnels.  Each near hall will host two 20 t
gadolinium doped liquid scintillator detectors, while the far hall
will host four such detectors. It is the most aggressive layout to
push systematics below 0.4\%.  Other experiments currently in data
taking exhibit some sensitivity to \thetaot.  They are MINOS, whose
measurement has already been mentioned and OPERA~\cite{OPERA} at the
Gran Sasso Laboratories (LNGS) of the Italian Institute for Nuclear
Research (INFN). In the original plans MINOS was supposed to improve
by a factor of two the limit of CHOOZ. However, considering the
achieved precision and the fact that MINOS has swapped to
antineutrinos, it is unlikely that it will be able to improve
significantly the current CHOOZ limit.  Also OPERA, which exploits the
above-mentioned CNGS beam, is expected to improve the CHOOZ
limit~\cite{opera_numunue}, although with a timescale that is not
competitive with T2K. On the other hand, in the occurrence of large
\thetaot (see below) the measurement from CNGS would be of great
value, being affected by very different systematics with respect to
T2K and Double-Chooz.

The observation of the extraordinarily large neutrino mixings when
compared to naive Cabibbo-like expectations from quarks, the
possibility of Majorana-like couplings and the extraordinary small
mass spectrum indicate that neutrinos might have unique features with
respect to other elementary fermions. This is why non-standard scenarios have
been deeply investigated in literatures, often fostered by
experimental anomalies as the one of LSND that dates back to
1995~\cite{lsnd} or, more recently, the ``low energy anomaly'' of
MiniBooNE~\cite{miniboone_lowe}.  Perspectives outside the standard
three-family framework are discussed in Sec~\ref{sec:nonstandard}.

\subsection*{Guessing the Future}
\begin{figure}
\centering
{\includegraphics[width=12cm]{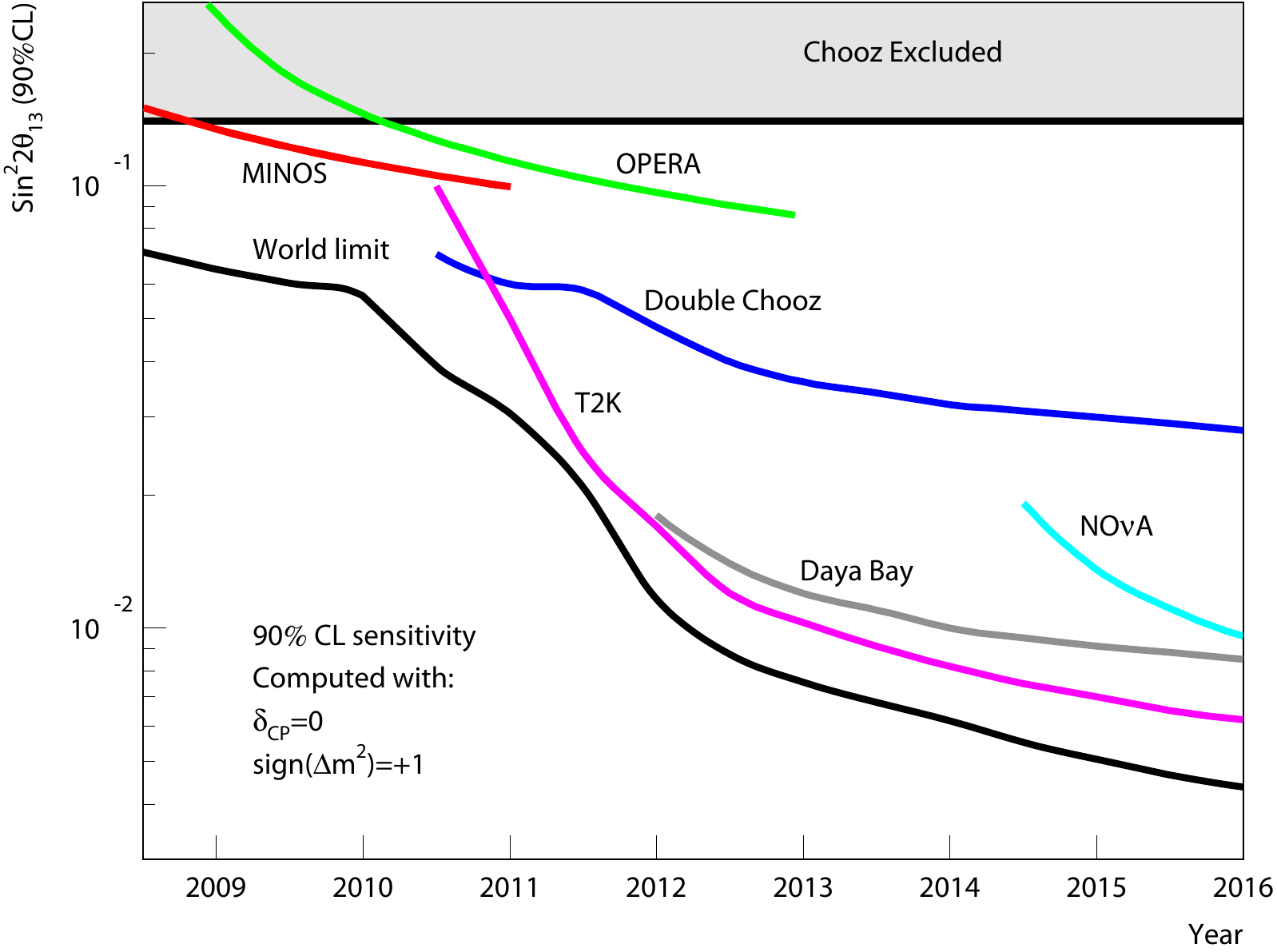}}
    \caption{Evolution of experimental $\sin^2{2\thetaot}$
    sensitivities as function of time.  All the sensitivities are
    taken from the proposals of the experiments.  For T2K it is
    assumed a beam power of 0.1 MW the first year, 0.75 MW from the
    third year and a linear transition in between.  NO$\nu$A
    sensitivity is computed for $6.5\times 10^{20}$ protons-on-target
    per year and 15000 ton detector mass.  Accelerator
    experiments sensitivities are computed for $\dmtt=2.5\times10^{-3}$
    eV$^2$, $\delta=0$ and normal hierarchy for all the experiments.
    The sensitivity curves are drawn starting after six months of data
    taking.  From \protect\cite{Mauro-th13}.  }
    \label{fig:th13vstime}
\end{figure}
It is of great practical interest to consider the expected
sensitivities of accelerator and reactor experiments in the near
future. It is, clearly, a quite approximate exercise since most of
these experiments are not in regular data taking and therefore, delays
or re-scheduling might be possible. This holds particularly for T2K,
whose startup will be at very low power (0.1 MW) compared with the
design one of 0.75~MW and a progressive ramp-up is foreseen.

\begin{figure}
\centering
{\includegraphics[width=12cm]{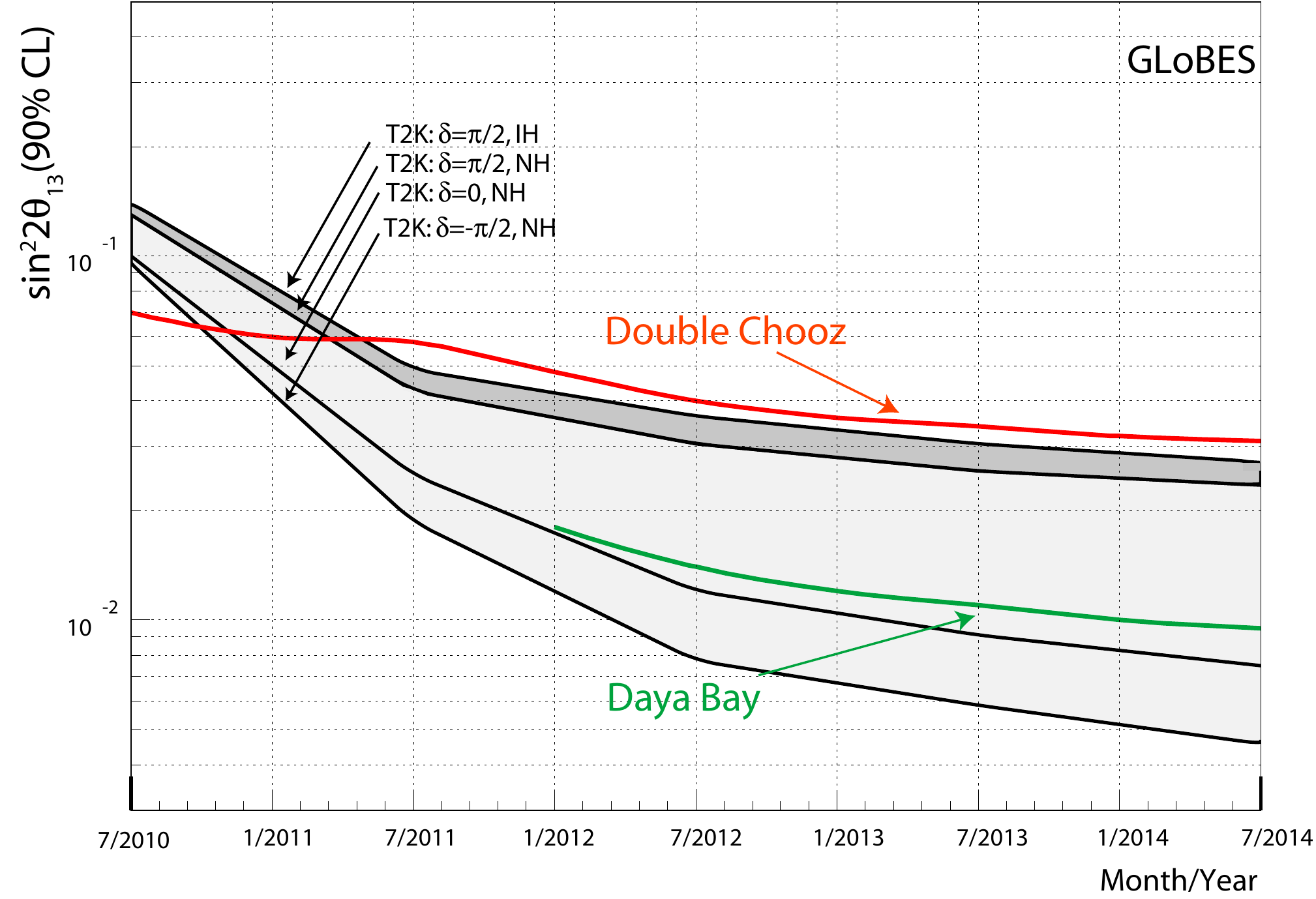}}
    \caption{Evolution of experimental $\sin^2{2\thetaot}$
    sensitivities of T2K and the reactor experiments Double-Chooz and
    Daya Bay as function of time, under the same assumptions of
    Fig.~\protect\ref{fig:th13vstime}.  The T2K sensitivity is
    computed with the Globes~\protect\cite{Globes} code and shown as a
    band of values evaluated ranging $\delta$ from 0 to $2\pi$ and
    assuming both normal (NH) and inverted (IH) hierarchy. From
    \protect\cite{Mauro-th13}.}
    \label{fig:time-evolution}
\end{figure}

Fig.~\ref{fig:th13vstime} \cite{Mauro-th13} shows the evolution of the
\thetaot sensitivities as a function of time gathering all the
information we have presently in our disposal\footnote{They
do not include the considerations on the final MINOS sensitivity
stated above.}.  Even with such caveat, from the plot one can easily
derive that in the next 5 years or so the \thetaot parameter will be
probed with a sensitivity about 25 times better than the present
limit.  The figure includes also the startup of NOVA~\cite{NOVA}, a
new facility in the US based on the existing NuMI beam from Fermilab
to Minnesota and its upgrade (see Sec.~\ref{sec:new_facility}).  In fact,
the \thetaot sensitivity of T2K depends on the unknown $\delta$
parameter and on the mass hierarchy, i.e. \sigdm; a more detailed
comparison of the time evolution of the T2K and reactors sensitivities
has to take into account that T2K will actually provide a band of excluded
values of \thetaot and not just a single point.  This is shown in
Fig.~\ref{fig:time-evolution}~\cite{Mauro-th13}.

From the plot some considerations can be taken:
\begin{itemize}
	\item
The Double-Chooz reactor experiment is very competitive in the first years of operation, when, however, the information
coming from its near detector will likely not be available.
	\item
T2K will be dominating our knowledge of \thetaot for long time and the
evolution of its beam power is the most critical issue in the
extrapolation of Fig.~\ref{fig:th13vstime}. It is impossible to state
now how the the JPARC neutrino beam line evolves and when it reaches
the top value of 0.75~MW, since the facility is based on an entirely
new accelerator complex. Therefore, the sensitivity shown here is no
more than an educated guess.
	\item
Scheduling is critical for Daya Bay, too. The achievement of tiny
systematic errors claimed by the experiment will likely be the most
relevant issue to be clarified.
\end{itemize}
This discussion is based on the limits that can be achieved by each experiment.
In case of \thetaot in the reach of those experiments, their information will be
truly complementary to measure the actual value of the parameter: for a discussion
under this hypothesis see \cite{Huber:2009cw,Lindner}.

\section{A new generation of facilities for the physics of
neutrino oscillations}
\label{sec:new_facility}

As already mentioned, there is consensus~\cite{Huber:2009cw} on the
fact that none of currently running facilities - including the
experiments in startup phase as T2K, Double-Chooz or Daya-Bay - will
be able to extract all missing parameters of the $U_{PMNS}$ even
in the most optimistic case of \thetaot close to current limits. On the
other hand, what has prompted a rather heated debate is the definition
of a ``minimal'' facility that might be able to address the challenge
of extracting such parameters by a detailed study of subdominant
transitions at the atmospheric scale. Defining the ``optimal
facility'' without knowing - at least approximately - the size of
\thetaot is a hopeless exercise but several non-trivial considerations
can be done assuming (or excluding) that an evidence of subdominant
\numunue transitions is within the reach of T2K.  Such considerations
have also driven important R\&D efforts toward novel sources of
neutrinos in the GeV range and, correspondingly, toward new detectors
for their observation at the far location.

The technologies for neutrino sources that are considered promising
for the next generation of facilities can be grouped in three main
streams~\cite{exotica} called, in jargon, ``Superbeams'',
``Beta~Beams'' and ``Neutrino Factories''~\cite{ISS_acc}. Coordinated
R\&D efforts have been made around the world to raise these
technologies to their adulthood but it is clear that a positive result
in the quest of \thetaot within the next 2-5 years will prompt a
tremendous boost toward their realization. The status of the art on
the eve of the T2K run is summarized below, with special emphasis to
their implementation in the European framework of
infrastructures.

\subsection{High Intensity neutrino beams from $\pi$ decays: the Superbeams}

\begin{figure}
\centering
{\includegraphics[width=10cm]{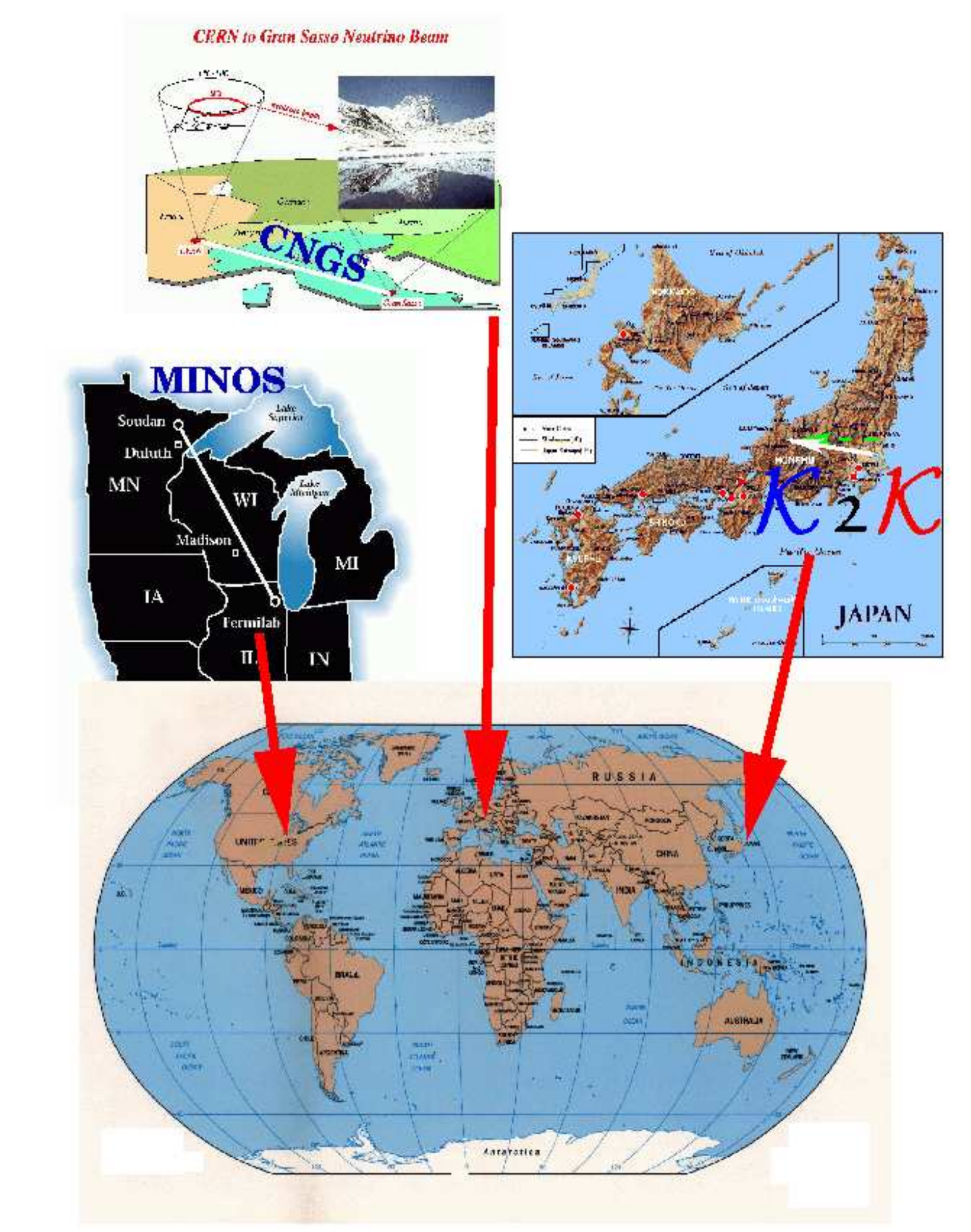}}
    \caption{The long baseline experiments run to gather evidence
for neutrino oscillations and measure the leading
parameters at the atmospheric scale (\dmtt\ and $\theta_{23})$. The
KEK-to-Kamioka (K2K) experiment in Japan from KEK to SuperKamiokande
(250~km) operated from 1999 to 2004; the MINOS experiment located
730~km far from Fermilab along the NuMI neutrino beam (still
running); the OPERA and ICARUS detectors located at the Gran Sasso
laboratories and detecting neutrinos from CNGS beam (still running). }
    \label{fig:longbaseline2}
\end{figure}

Since the 1962 experiment of Lederman et al.~\cite{steinberger},
accelerator neutrinos in the GeV ballpark have only been produced from
the decay in flight of pions. This 40-year old technology~\cite{kopp}
has progressed together with high intensity proton accelerators and,
in the last decade, allowed for the construction of long-baseline
$\nu_\mu$ beams to explore the leading \numunutau\ transition at the
atmospheric scale (Fig.~\ref{fig:longbaseline2}). Long baseline
experiments have also been possible thanks to the large size of the
atmospheric mixing ($\sin^2 2\theta_{23} \simeq 1$), which partially
compensates the loss in intensity due to the large source-detector
distance. At CERN, it allowed for the re-use of the SPS, formerly
employed for short-baseline experiments at the WANF
(CHORUS~\cite{Eskut:1997ar} and NOMAD~\cite{Astier:2003gs}) to feed a
17 GeV beam from CERN to Gran Sasso.

 Moving from leading to subleading oscillations (\numunue
transitions), the bonus of the large mixing angle disappears (\thetaot
$\ll$ \thetatt) and the intensity is no more sufficient to explore
these transitions at a level better than CHOOZ. Such considerations
hold also for other long-baseline beams in the world that played a
role in the current round of experiments, as the already mentioned
NuMI beam from Fermilab to Minnesota and the K2K beam from KEK to
SuperKamiokande (see Fig.~\ref{fig:longbaseline2}). In Japan, it would
have been impossible to re-use the very low intensity K2K beam in this
new phase. As a result, the construction of a new facility has been
pursued since 2003, leveraging the enormous investment (about 1
Billion\textgreek{\euro}) made for JPARC and aimed to a much broader
community than neutrino physicists. It brought the T2K project into
existence without the need of building either a dedicated proton
accelerator or a new far detector: the former leverages the 50~GeV
general purpose synchrotron of JPARC complemented by a dedicated neutrino
beamline, the latter the existing SuperKamiokande detector;  although
the additional investment for the neutrino beamline was
demanding (about 200 M\$), it comes to no surprise that the T2K
project is by far the most advanced among the experiments conceived to
improve the current limits of \thetaot. Its running schedule has
already been discussed in Sec.~\ref{sec:intro}.

\begin{figure}
\centering
{\includegraphics[width=12cm]{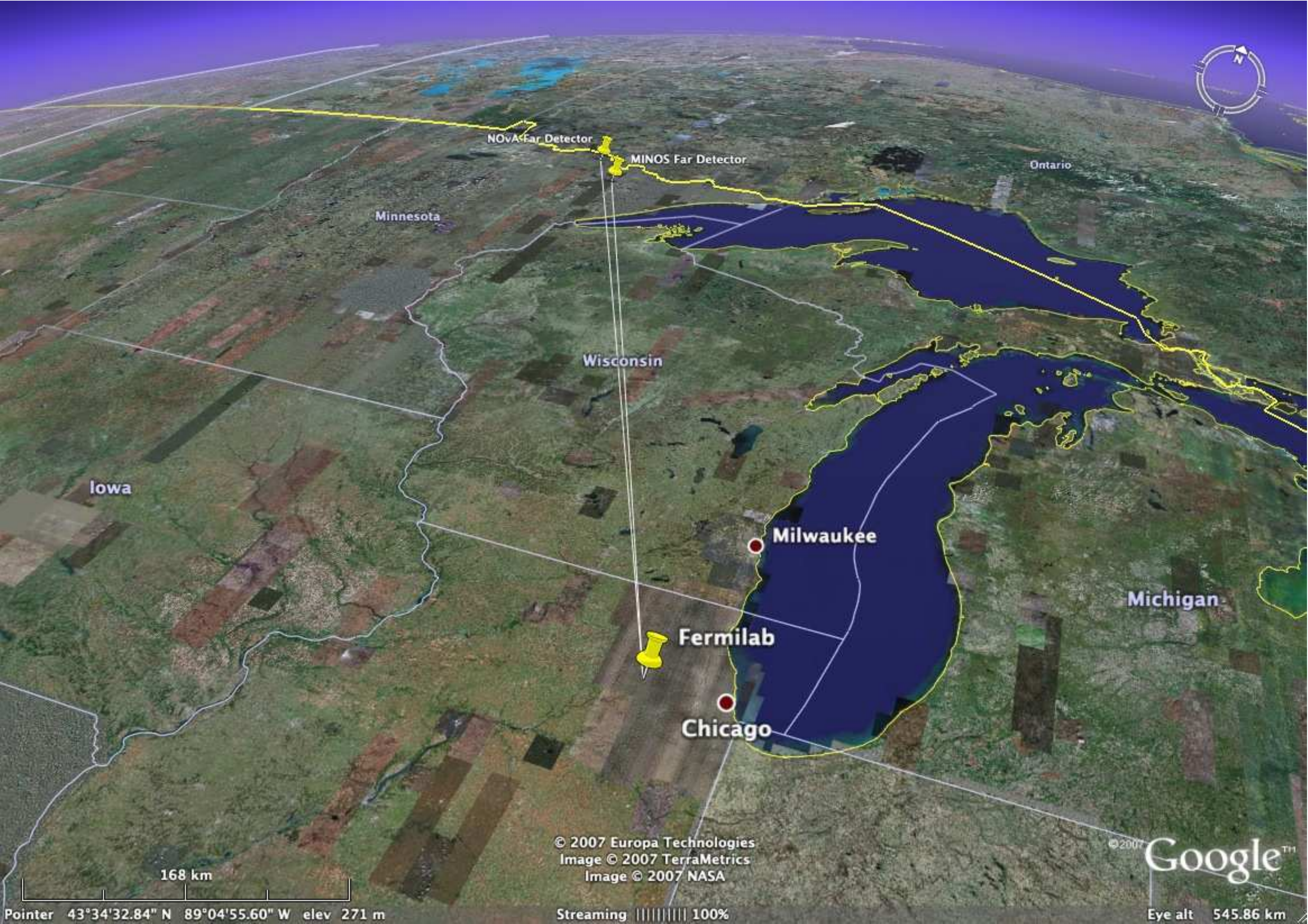}}
    \caption{The location of the MINOS experiment along the axis of
    the NuMI beam in the US and the location of the new NOVA
    detector, exploiting the same beam off the main axis. }
    \label{fig:nova}
\end{figure}

\begin{figure}
\centering
{\includegraphics[width=15cm]{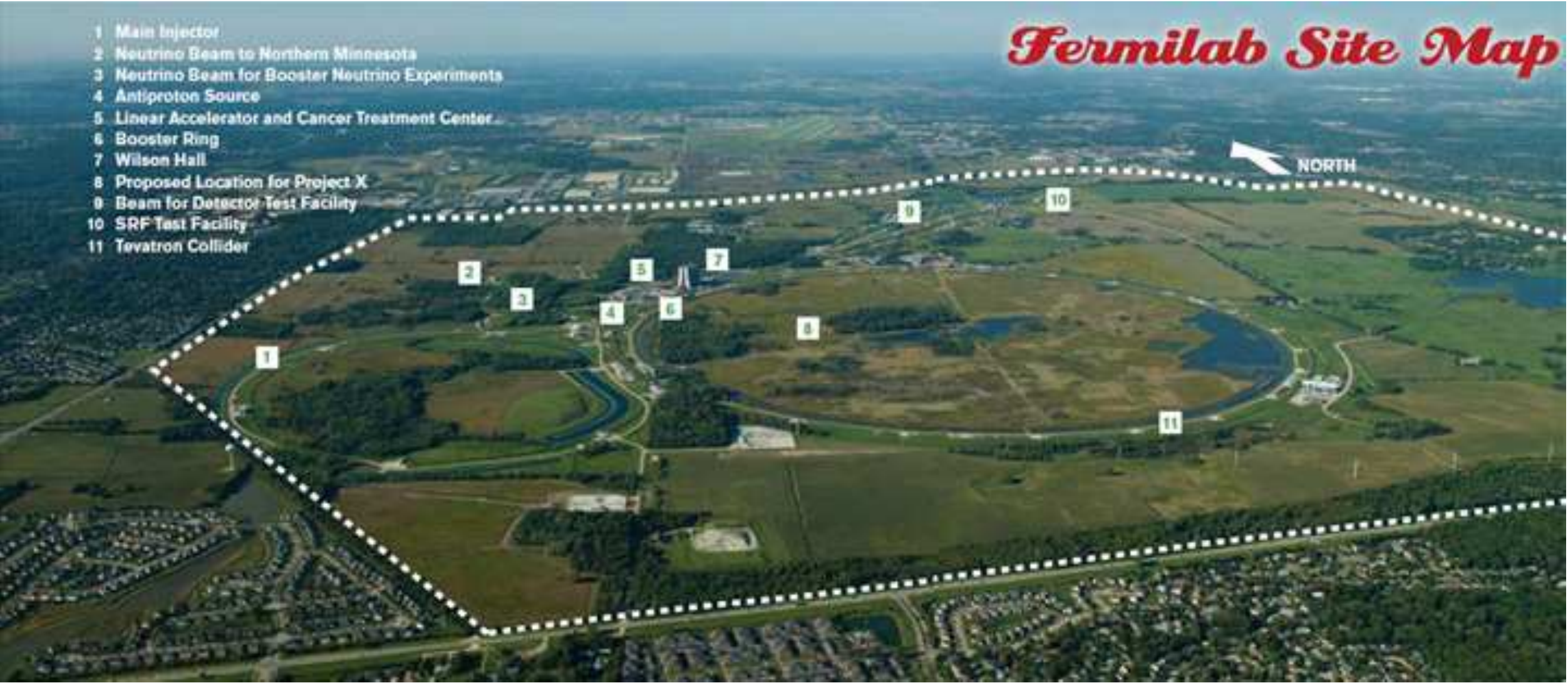}}
    \caption{The Fermilab site and the location of the Main Injector
(1) and Tevatron (11) together with the proposed location of Project X
(8).}
    \label{fig:fermilab}
\end{figure}

 Both US and Europe have neutrino facilities much more powerful than
the K2K beamline. The NuMI beam might be operated up to 0.4 MW,
i.e. at 40\% of the maximum power achievable by T2K.  Unfortunately,
there is no far detector in the US that can exploit such power since
MINOS is too dense to perform precision measurements of \nue
appearance. The construction of a dedicated 15~kton detector
(NOVA~\cite{NOVA}), started on May 2009 and located 810~km from Fermilab (see
Fig.\ref{fig:nova}) brings the timeline of the first physics results a
few years after T2K and without significant improvements on the
knowledge of \thetaot but with some marginal sensitivity to the mass
hierarchy due to the large baseline (810 versus 292~km). However, the
existence of NOVA together with a positive result from T2K could
ground a second generation facility, based on a new neutrino beam fed
by a 2 MW, 8 GeV, proton accelerator (``Project
X''~\cite{projectX}). In fact, as a part of the NOVA project, Fermilab
plans an upgrade of NuMI based on the exploitation of the Fermilab
``Recycler Ring'' after the shut-down of the Tevatron. The Recycler
Ring will be used as a pre-injector of the Fermilab ``Main Injector''
reducing the cycle time (Fig.\ref{fig:fermilab}). It needs, however,
the construction of a dedicated transfer line and the upgrade of the
radio-frequency system in both the Recycler and in the Main
Injector. This upgrade~\cite{ANU} could finally bring NuMI to an peak
power comparable to T2K: 0.7 MW and $6 \times 10^{20}$
protons-on-target per year (pot/y) at 120 GeV.

\begin{figure}
\centering
{\includegraphics[width=12cm]{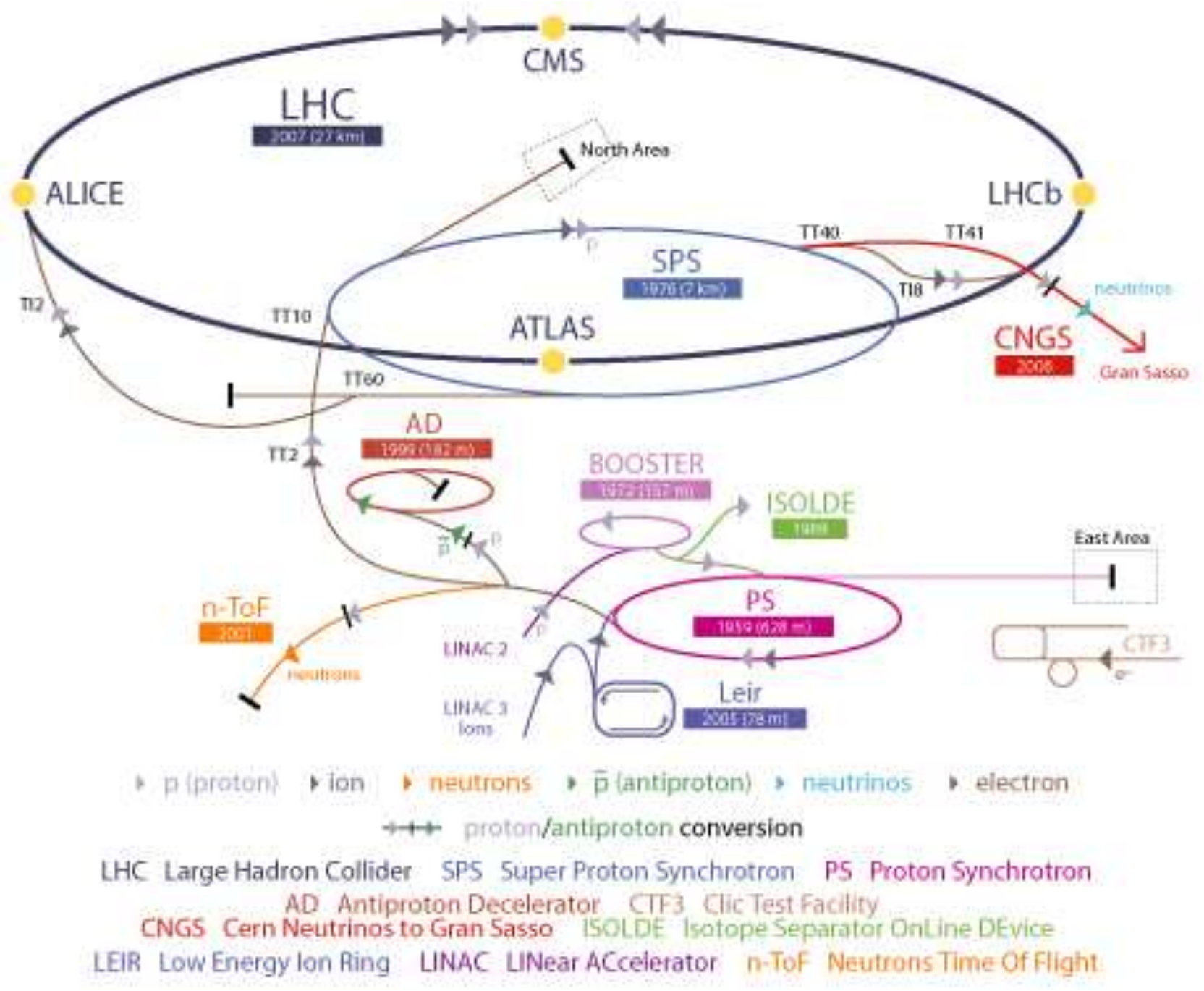}}
    \caption{The CERN accelerator complex, which includes the
Proton Synchrotron (PS), the Super-Proton-Synchrotron (SPS), the Large
Hadron Collider (LHC) and the CERN Neutrino beam to Gran Sasso
(CNGS).}
    \label{fig:cern}
\end{figure}

In Europe, the perspectives for a high intensity neutrino experiment
based on Superbeams are entangled with the evolution of the CERN
acceleration complex and, in particular, of the injection system of
the LHC.  The CNGS at nominal intensity can be operated to accumulate
$4.5 \times 10^{19}$ pot/y at an energy\footnote{To ease comparison
with NOVA and following~\cite{modular}, we suggest, as a rule of
thumb, to scale the integrated pot with the linear ratio of the
primary proton energy. Hence, a nominal CNGS year corresponds to $1.5 \times
10^{20}$ ``NOVA'' pot/y.}  of 400~GeV.  In the last few years,
particularly in the framework of the CERN PAF (``Proton Accelerators
for Future'') Working Group, it has been investigated~\cite{meddahi}
the possibility of increasing the intensity of the CNGS both using
present facilities and, on a longer timescale, exploiting an upgrade
of the acceleration complex (Fig.\ref{fig:cern}). The ultimate CNGS
performance are actually limited by the injection from the 50-year-old
Proton Synchrotron (PS). In this scenario (CNGS as the only user of
the SPS at CERN beyond the LHC), the facility could deliver up to $1
\times 10^{20}$ pot/y ($3.3 \times 10^{20}$ NOVA pot/y). At a longer
timescale ($>$2016), the replacement of the PS with a new 50 GeV
synchrotron (PS2~\cite{PS2}) might surpass these limitations, provided
an appropriate upgrade of the SPS radio-frequency system. It would
bring CNGS to a maximum intensity (CNGS as only user of the SPS beyond
the LHC) of $2 \times 10^{20}$ pot/y ($6.6 \times 10^{20}$ NOVA
pot/y). It is, therefore, a situation that closely resembles the one
of the NuMI upgrade but unfortunately, its realization cannot be
anticipated before 2017 due to the priority to the running of the LHC
and the corresponding funding limitations from CERN for the
construction of the new injectors. At that timescale, it would be more
interesting to consider a high-flier option, specifically suited for
the exploration of CP violation and the mass hierarchy in case of
positive result from T2K - along the line of the US ``Project X''.
The rationale of this approach is that a multi-MW proton source can,
in principle, serve for a variety of needs and feed a rich physics
programme ranging from neutrino physics to muon and kaon physics and
to physics with radioactive ion beams~\cite{bettoni,marciano}. This
option has been considered at CERN in the framework of the
construction of a pre-injector for PS2: the Superconducting Proton
Linac (SPL).  The pre-injector, feeding PS2 with 2-4~GeV protons, has
been conceived in year 2000 as a 2.2~GeV accelerator for a future
neutrino factory based at CERN and its design was based on the
recycling of the radio-frequencies formerly employed at
LEP~\cite{SPL-Design}. Over the time, the design has been updated and
currently points toward a 3.5-4 GeV, 4 MW system with dedicated
equipment based on bulk Niobium cavities~\cite{SPL-II}. This proton
energy might feed a Superbeam producing neutrinos at a rather low
energy (about 300 MeV) that could be exploited by a very massive
detector located not far from CERN (CERN-to-Frejus~\cite{memphis}).

The technology of the Superbeam is considered the most straightforward
way to explore leptonic mixing in case of positive result from T2K
and/or Double-Chooz and, therefore, it is viewed as a solid option in
Japan (Tokai to HyperKamiokande - see below) and US (Fermilab/Project X to
DUSEL)~\cite{ISS_phys}.  Unfortunately, it also shows
evident limitations:
\begin{itemize}
\item It is not a ``pure'' source of neutrinos of a given flavor,
being plagued by the $\nu_e$ produced by the decay-in-flight of the
kaons and of the muons. When seeking for subdominant \numunue
transitions, the systematics on the knowledge of the $\nu_e$
contamination will likely be  the main limitation for a precise
determination of CP violation in the leptonic
sector~\cite{Huber:2007em};
\item It produces mainly $\nu_\mu$ and, therefore, the leptonic mixing
is studied through \numunue and its CP conjugate \nubarmunubare, i.e.
looking for {\it electrons} in the final state instead of muons. It
thus requires huge, low density detectors to be hosted in underground
sites (see below);
\item in a Superbeam, CP violation appears as an asymmetry between the
probability of transition \numunue and its CP conjugate
\nubarmunubare.  At low proton energies, the production yield of
$\pi^-$ is suppressed with respect to $\pi^+$. This suppression,
together with the suppression due to the cross section ($\sigma_{\bar{\nu}} /
\sigma_\nu \simeq 1/2$), makes the antineutrino run much more
time-consuming that the neutrino run.
\end{itemize}

\noindent
A multi-MW proton accelerator and the corresponding neutrino beamline is
considered feasible in less than a decade~\cite{Hylen} but it requires a
substantial effort in accelerators and targetry, a major resource
investment (of the order of 500 M\textgreek{\euro}) and feedback from the operation
of neutrino facilities in the sub-MW range (firstly T2K).

\vspace{0.5cm}
\noindent
{\bf Far Detectors for the Superbeams}
\vspace{0.5cm}

\noindent
Most of the information gathered at the atmospheric scale on leptonic
mixing come from the study of muon neutrino disappearance in natural
(atmospheric) and artificial (accelerator) sources. The use of
Superbeam forces experimentalists to build detectors that are
specifically designed for the identification of electron neutrinos.
$\nue$ appearance is extremely challenging compared with $\numu$
disappearance because prompt electromagnetic showers must be
distinguished from neutral-currents events where electromagnetic
showers occur through the production of $\pi^0$. A low density, high
granularity target is mandatory to allow showers to develop and to
perform a precise kinematic reconstruction in order to separate
effectively neutral-current (NC) from charged-current (CC)
events. Unfortunately, only a few of the techniques currently applied
for electromagnetic calorimetry can be scaled to a size suitable for
neutrino oscillation physics (at least tens of ktons). Actually, due
to the current constraints to the size of \thetaot, it is believed
that even in the occurrence of a clear evidence for \numunue
oscillations at T2K, the size of the detectors should be upgraded by
nearly an order of magnitude with respect to the current generation of
experiments.  Thanks to the experience of SuperKamiokande, Japan
maintains a leadership in Water Cherenkov detectors and the proposal
of building a new very massive detector of about 500 kton fiducial
mass is considered the most plausible upgrade of T2K (the
``megaton-size'' Hyper-Kamiokande detector~\cite{HK}), possibly
complemented with a second detector in
Korea~\cite{Kajita:2008zzc}. Similar options
are considered in US for the new DUSEL underground laboratory in South
Dakota, and in Europe for a possible major extension of the Frejus
Labs.  The only two other viable technologies that are considered
scalable to such size~\cite{Autiero:2007zj,ISS_det} are the ones of
the Liquid Argon detectors and of liquid/plastic
scintillators. C.~Rubbia proposed the use of liquid Argon Time
Projection Chambers (LAr TPC) since 1977~\cite{rubbia} and INFN
maintains a world leadership in the LAr technology thanks to the
results of the ICARUS Collaboration that built such detectors up to a
size of 600~tons~\cite{Amerio:2004ze}. Due to its finer granularity
and superior quality in reconstructing and identifying particles and
rejecting NC events, LAr can achieve sensitivities to $\nu_e$
appearance comparable or better than SuperKamiokande with smaller
masses (in the 10~kton range). To catch up with the megaton-size Water
Cherenkov detectors, LAr TPC's should reach fiducial volumes of about
100~kton: an extrapolation of about two order of magnitudes with
respect to the T600, which necessarily requires a very intensive R\&D.
This phase might be substantially reduced following a modular
approach, where the 10-20 kton initial mass can be successively
enlarged as needed.

\vspace{0.5cm}
\noindent
{\bf Synergies with non-accelerator astroparticle physics}
\vspace{0.5cm}

Bringing liquid detectors to this immense size requires experimental
halls that do not exist in any part of the world. If the technology of
Superbeams is chosen as a reference for the post-T2K phase, the Gran
Sasso Laboratories will not be able to host any far detector.  In
this framework, an intermediate size (20~kton) detector in Italy could
only be hosted by a new shallow-depth laboratory, as proposed
in~\cite{modular}.  On the other hand, large liquid detectors with a
fiducial mass significantly bigger than SuperKamiokande come with a
bonus of a physics programme much broader than neutrino
oscillations~\cite{Autiero:2007zj} and, therefore, are at the focus of
important R\&D programmes worldwide.  If located underground or - at
least for LAr - even at a relatively shallow depths ($\sim 500$
m.w.e.~\cite{Bueno:2007um}), they could improve significantly the
current measurements on proton decay, supernovae, solar and
atmospheric neutrinos and, hence, they would represent a major step forward in
experimental astroparticle physics.

In 2007 the European Union has approved a Design Study
(LAGUNA~\cite{LAGUNA}) aimed at evaluating extensions of the
present deep underground laboratories in Europe (Boulby, Canfranc and
Frejus) in order to host such facilities; it also considers the creation of new
laboratories in the following regions: Caso (Italy), Pyhasalmi
(Finland), Sierozsowice (Poland) and Slanic (Romania). The LAGUNA DS
is mainly oriented to infrastructure studies for the underground sites
since these new halls pose major engineering problems. In parallel,
the physics case and the technological challenges for the detectors to
be hosted are pursued by most European countries following three
baseline technologies: water Cherenkov (the 500~kton mass
MEMPHIS~\cite{memphis,memphis_det} detector), liquid scintillator (the 50~kton
LENA~\cite{LENA} detector) and liquid Argon TPC's (the 100 kton
GLACIER~\cite{GLACIER} LAr TPC). Clearly, the choice of an European
laboratory as a possible hosting site would boost enormously the
Superbeam option although (see below) the Beta Beams could profit of
this unique opportunity, too.

\subsection{Neutrinos from $\mu$ decays: the Neutrino Factory}
\label{sec:NF}

Production, acceleration and stacking of high intensity muon beams for
muon colliders have been envisaged since the 60's and it has been
noted very early that their decays might produce useful beams of
$\numu$ and $\nubare$ (exploiting $\mu^-$ decays into $e^- \nubare
\numu$) or $\nubarmu$ and $\nue$ ($\mu^+$ decays into $e^+ \nue
\nubarmu$). However, realistic layouts to get intense neutrino sources
have become available only in recent times~\cite{NF_history}. In the
modern formulation of the ``Neutrino Factory'' concept, muons are
created from an intense pion source at low energies, their phase space
compressed to produce a bright beam, which is then accelerated to the
desired energy and injected into a storage ring with long straight
sections pointing in the desired direction. In 1997
S.~Geer~\cite{geer:1997} noted that this source could be ideal to
study \nuenumu\ oscillations at the atmospheric scale, i.e. the
T-conjugate of the channel observed in Superbeams (\numunue).  Since
$\mu^+$ decay into $e^+ \nue \nubarmu$, it is possible to investigate
\nuenumu\ oscillations seeking for the appearance of $\mu^-$ from
$\numu$ CC events (``wrong sign muons''), provided that we are able to
separate these events from the bulk of $\mu^+$ (``right sign muons'')
coming from unoscillated $\anumu$. A. De Rujula et al.~\cite{De
Rujula:1998hd} underlined that the simultaneous exploitation of
$\mu^-$ and $\mu^+$ decays would be an ideal tool to address CP
violation in the leptonic sector, with outstanding performance
compared with pion-based sources. The only condition to be fulfilled
was a large ratio \dmot/\dmtt (see Sec.~\ref{sec:intro}) and, of
course, the finite size of \thetaot. The results from KAMLAND and
SNO~\cite{SNO} confirming $\dmot/\dmtt \simeq 1/30$ boosted enormously
Geer's proposal, together with the above-mentioned proposals for
Superbeams. Moreover, Geer's ideas resonated with the needs of the
Muon Collider accelerator community, who appreciated the possibility
of a strong physics-motivated intermediate step before facing the
enterprise of the Muon Collider itself. Actually, motivations were so
strong that the early ideas of building the Neutrino Factory (NF) as a
front-end of the muon collider have been dropped and current designs
are all optimized for neutrino physics. It mainly implied a relaxation
of the ultra-challenging constraints on muon cooling needed for the
construction of the collider~\cite{blondel_nufact09}. In spite of
this, the realization of the neutrino factory still represents a major
accelerator challenge compared with Superbeams. It is met through a
world-wide R\&D programme~\cite{bross}; in Europe this programme is
especially fostered by UK. Among the NF-oriented projects we recall
MICE~\cite{MICE} at the Rutherford Appleton Laboratories (ionization
cooling), HARP~\cite{HARP} at CERN (hadroproduction for the front-end
proton accelerator), MERIT~\cite{MERIT} at CERN (targetry),
EMMA~\cite{EMMA} at Daresbury (fixed-field alternating-gradient
accelerators) and the MUCOOL R\&D~\cite{MUCOOL} at Fermilab
(radio-frequency and absorbers).  Moreover, the NF has to be seeded by
a very powerful low-energy proton accelerator (4~MW); its realization
requires similar R\&D as for the Superbeams, although its optimal
energy lays in the few-GeV range (e.g. the aforementioned SPL).
Current designs aim at 10$^{21}$ muon decays per year running with a
muon energy of 50~GeV, although more realistic scenarios suggest decay
rates in the ballpark of $2-5 \times 10^{20}$ decay/y and with
energies in the 20-50~GeV range (corresponding to neutrinos in the
10-30~GeV range)~\cite{review-geer}. After the work of the
International Scoping Study (ISS)~\cite{ISS_acc,ISS_phys,ISS_det},
there is a rather widespread consensus on the fact that the Neutrino
Factory can be considered the most performing facility for the
determination of \thetaot, CP violation and the mass hierarchy. With
respect to Superbeams, they profit of much smaller systematics in the
knowledge of the source and much higher energies (i.e. statistics, due
to the linear rise of the deep-inelastic $\nu_\mu$ cross section with
energy). In fact, the energy is so high that for any realistic
baseline ($<7000$~km) the ratio L/E will be off the peak of the
oscillation maximum\footnote{The peak of the oscillation occurs (see
Eq.\ref{eq:pontecorvo}) when $1.27 \dmtt L/E = \pi /2$. This situation
holds also at CNGS, where OPERA and ICARUS are located at 730~km but
the neutrino energy cannot be lowered below {\cal O}(10) GeV to stay
beyond the kinematic threshold for tau production. In this case, the
oscillation peak is at $1.27 \cdot 2.5\times 10^{-3} (\ \mathrm{eV}^2)
\ \cdot 730 \ \mathrm{(km)} \cdot 2/\pi \ = \ 1.5 \
\mathrm{GeV}$. Similarly, at a Neutrino Factory with a detector
located 3000~km far from the source, the peak is at 6~GeV.}  at the
atmospheric scale. This condition is the main cause of the occurrence
of multiple solutions when the mixing parameters are extracted from
the physics observables, i.e. the rates of appearance of wrong sign
muons. In jargon, this issue is dubbed the ``degeneracy
problem''~\cite{Barger:2001yr,ISS_phys} and it has already been
mentioned in Sec.~\ref{sec:intro}. It also affects other facilities
than NF but it is particularly severe for experiments running off the
peak of the oscillation probability. The ISS suggests as an ideal
solution the positioning of two detectors at baseline around 3000 and
7000~km.  An alternative to the second 7000~km detector could be the
detection of \nuenutau\ at baseline around 1000~km (``silver
channel'')~\cite{silver} (see below).  The exploitation of the silver
channel, moreover, is useful to investigate the occurrence of
non-standard interactions in the neutrino
sector~\cite{kopp_j}. Although the superior physics reach of the
Neutrino Factory is nearly undisputed and no evident showstoppers have
been identified, the R\&D needed to build this facility remains
impressive. In turn, the time schedule for its realization and the
cost estimate are vague ($\sim 2020$ after an investment of 1-2
Billion\$). On the other hand, a clear indication on the size of
\thetaot will enormously boost the interest of particle physics on
this technology. Neutrino Factories are virtually capable of
performing real precision physics on the leptonic mixing in a way that
resembles the former physics potential of the b-factories on quark
mixing.

\vspace{0.5cm}
\noindent
{\bf Far Detectors for the Neutrino Factories}
\vspace{0.5cm}

\noindent
At the Neutrino Factory, experimentalists search for the appearance of
multi-GeV $\nu_\mu$, a signal with a very clear topology even in high
density, low granularity detectors. As a result, the baseline
technology for NF far detectors are magnetized iron calorimeters with
masses of the order of several tens of ktons, although other
low-density options have also been investigated. Magnetized iron
calorimetry is a classic technique in neutrino physics, brought to the
multi-kton size by the MINOS Collaboration. Similar techniques based
on Resistive Plate Chambers instead of plastic scintillators have been
proposed in the past (MONOLITH~\cite{MONOLITH}) for a second
generation atmospheric neutrino experiment in the Hall C of LNGS - the
same Hall that presently hosts OPERA and Borexino~\cite{borexino} -
and, more recently, in a new underground lab close to Bangalore in
India (INO~\cite{INO}). They have been proposed to measure the
sinusoidal patter of the neutrino oscillations with multi-GeV
$\nu_\mu$ and $\anumu$ produced in the earth atmosphere, improve the
knowledge of the leading atmospheric parameters $\thetatt$ and $\dmtt$
and, at least for large values of $\thetaot$ ($ > 5^\circ
$~\cite{samanta}), determine the sign of \dmtt\ through matter
effects.  The main challenge with respect to MINOS is represented by
the compelling request on the muon charge identification ($\ll
10^{-3}$ of misidentification probability) in order to suppress the
bulk of ``right sign muons''. Current R\&D are focused on the
optimization of the granularity and on the technology choice for the
active detector to achieve optimal charge identification and pion
rejection at the lowest possible visible energy~\cite{MIND}. Clearly,
Hall C of LNGS is an ideal place to host a 30-40~kton
detector\footnote{In the original MONOLITH proposal, a 30 kton
detector had to be installed next to Borexino, in the area presently
filled by the OPERA detector. A significantly larger mass could be
achieved in future.}. A Neutrino Factory based at the Rutherford Labs (RAL), along
the line pursued by UK, could be suited for a LNGS far-detector,
although the RAL-LNGS distance is about 1500~km (i.e. a factor of two
smaller than the optimal one suggested by the ISS). In the context of
a CERN-based NF, LNGS is too close for an optimal sensitivity to CP
violation or mass hierarchy. On the other hand, the LNGS location is
interesting to study the ``silver channel'' (\nuenutau). The silver
channel represents a unique opportunity to exploit technologies
mastered within INFN~\cite{emulsion}: the appearance of the \nutau\
can be observed by nuclear emulsion based detectors in a way similar
to the one currently employed by OPERA at CNGS.  In the Neutrino
Factory framework, the detectors should be suited for the
identification of the tau lepton, provided that they are able to
measure the wrong-sign muons. An OPERA-like detector~\cite{silver} in
the 5~kton mass ballpark or a LAr TPC with comparable mass
complemented by a muon spectrometer are considered as viable options.
The only drawback might be due to the early construction of the Indian
laboratory, which is located at about 7000~km both from RAL and from
CERN. If this detector is built, the use of the ``silver channel'' as
a way to solve the degeneracies in a European NF will be marginal
compared with the baseline ISS scenario (two iron detectors at 3000
and 7000 km, respectively).

\subsection{Neutrinos from radioactive ion decays: the Betabeam}

The enormous progress in the technology of Radioactive Ion Beams has
led P. Zucchelli~\cite{BetaBeam} to the proposal of a neutrino
facility based on the decay in flight of $\beta$-unstable ions (for a
full review see \cite{BB-book}). Unlike the NF, these ``Betabeams''
(BB) are pure sources of \nubare \, or, in the occurrence of $\beta^+$
decays, of \nue. Hence, they are ideal tools to study \nuenumu\
transitions and their CP-conjugate. They share with NF the nearly
complete absence of systematics in the knowledge of the source with
the bonus of no ``right sign muon'' background (no $\numu$ in the
initial state). On the other hand, due to the very different
mass-to-charge ratio between muons and $\beta$-unstable ions, the
energy of the neutrinos are typically much smaller than what can be
obtained at the NF. The original proposal of \cite{BetaBeam} was tuned
to leverage at most the present facilities of CERN - the PS and the
SPS - and it was based on \He\ and \Ne\ as \nubare \, and \nue \,
sources respectively. It goes without saying that the Betabeam
triggered the interest of nuclear physics community, which was offered
a stimulating synergy with the neutrino programme at CERN. As a
result, such proposal~\cite{memphis,performance_BB} was studied in a
systematic manner within the framework of the EURISOL Design
Study\footnote{The EURISOL Design Study was a Project funded by the
European Community within the 6th Framework Programme as a Research
Infrastructures Action under the "Structuring the European Research
Area Specific Programme". The Project started officially on Feb 2005,
and has been completed on spring 2009.}  (Task 12: Beta Beam
aspects). The study aimed at $2.9 \times 10^{18}$ antineutrinos per
year from \He\ and $1.1 \times 10^{18}$ neutrinos per year from
\Ne. The outcome was extremely encouraging, except for the production
of \Ne, which cannot attain the needed rate using standard methods and
medium-intensity proton accelerators (200~kW). Along this line, the
most straightforward alternative would be direct production on MgO
based on a 2~MW, few MeV, proton accelerators, which are quite similar
to the linacs that have to be built for the International Fusion
Materials Irradiation Facility~\cite{IFMIF}. In this case, the BB
would partially miss the advantage of a low-power front-end compared
with the multi-MW accelerators needed for the Superbeams and for the
NF, although a few tens of MeV MW accelerator is anyway a much simpler
machine than a few GeV MW Linac.  From the point of view of the
physics performance, an additional weakness stems from the fact that
the SPS is able to accelerate these ions up to relatively low
energies, so that the corresponding emitted neutrinos are just in the
sub-GeV range. This impacts on the choice of the detectors (see below)
and on the cross sections, which are highly depleted.

 To improve the performance of the Eurisol Beta Beam several
alternatives to the SPS have been considered: a high energy SPS
(``SuperSPS''~\cite{SuperSPS}) accelerating protons up to 1~TeV (a machine
originally envisaged for the energy and luminosity upgrade of the LHC)
or even the LHC itself~\cite{highenergy_BB,LNGS_BB}.  These
configurations improve the sensitivities to CP violation and the mass
hierarchy at the expense of a large increase of costs: large
investments are needed especially to the construction of the decay
ring since the length of the ring depends from the magnetic rigidity
of the circulating ions, which is proportional to their Lorentz
$\gamma$ factor, and for the compensation of potential flux reduction
due to the longer lifetime of the ion in the laboratory frame.

In 2006, C.~Rubbia et al.~\cite{Rubbia:2006pi} proposed the use of
\Li8 and \B8 as neutrino sources noting that these isotopes could be
produced in a multiturn passage of a low-energy ion beam through a low-Z
target. In this case, ionization cooling techniques could increase the
circulating beam lifetime and thus enhance the ion production to a
level suitable for the Betabeam. This option has the advantage of
employing isotopes with higher Q-value than \Ne\ and \He, increasing
correspondingly the neutrino energy (from $\sim0.5$ to $\sim1.5$~GeV
for the SPS-based BB). This alternative approach will be at focus in
the framework of the EURO$\nu$ Design
Study\footnote{EURO$\nu$~\cite{euronu} is a FP7 Design Study which
started on Sept 2008 and will run for 4 years. The primary aims are to
study three possible future neutrino oscillation facilities for Europe
(a Superbeam from CERN-to-Frejus, a RAL or CERN based NF and high-Q
BB) and do a cost and performance comparison.}. A drawback with
respect to the use of low-Q ions is that the flux at the far location
is smaller due to the larger beam divergence and a larger amount of
ions stacked in the decay ring is needed. Although the BB optimization
is a complex task~\cite{Donini:2008zz, Agarwalla:2008gf,
Winter:2008dj}, some simple scaling laws can be used as reference. For
a given number of decays per year $N_\beta$, if we label with $\gamma$
the Lorentz factor of the ion (which depends on the machine employed
to accelerate the ion), with $Q$ the Q-value of the isotope and with
$L$ the source-detector distance, the events at the far location are
proportional to the convolution of the flux ($\phi \sim
\gamma^2/L^2$), of the cross section ($\sigma \sim E_\nu \sim Q
\gamma$) and of the oscillation probability, times $N_\beta$. If the
facility is operated at the maximum of the oscillation probability,
then $1.27 \Delta m^2 L/E_\nu= \pi/2$; therefore, $L \sim Q
\gamma$. As a result the number of events are proportional to $N_\beta
\gamma/Q$. Note, however, that also $N_\beta$ has a dependence on
$\gamma^{-1}$ due to the increase of the ion lifetime in the lab frame
at larger $\gamma$.

Summarizing, a high-Q BB needs a smaller $\gamma$ for the same
neutrino energy, with the advantage that an accelerator of larger
energy than the SPS would not be needed and that the length of the
decay ring could be shortened. On the other hand, given the Beta Beam
kinematics, for the same baseline L an high-Q BB needs an order of
magnitude more ions at the source to match the performance of a
high-$\gamma$ BB.

In general, the
clarification of the issue of the ion production yield is considered a
crucial milestone for the Betabeam. Given an appropriate yield, the
acceleration and stacking is viewed as less demanding than what is
needed for a NF both from the point of view of R\&D and cost. Clearly,
the possibility of employing existing facilities (e.g. the CERN PS-SPS
complex or its upgrades) might substantially strengthen this option.

\vspace{0.5cm}
\noindent
{\bf Detectors and experimental challenges for the Betabeams}
\vspace{0.5cm}

\noindent
A far detector within a BB facility seeks for the appearance of
$\numu$ (or \nubarmu \, during the run with \He\ and \Li8) in a bulk of
unoscillated \nue. Therefore, their detection can be even simpler than
for the NF, charge identification being immaterial in this case.  In
the SPS-based BB option, the energy of the outcoming neutrinos lays in
the sub-GeV range. Here, the range of the muons is comparable with the
one of the pions and high density detectors cannot perform a clear
NC/CC separation. Atmospheric neutrinos in such range are typically
studied with water Cherenkov detectors or with LAr TPC's. Therefore,
the ideal technology for such ``low-energy'' BB turns out to be
identical to the ideal technology for the Superbeams. Moreover, even
for nominal ion fluxes, the smallness of the cross section requires
very large detector masses, again in the range of the Superbeam far
detectors (e.g. the above-mentioned Hyper-Kamiokande
detector). Clearly, with respect to a Superbeam, the BB offers a
nearly complete control of the source systematics at the expenses of a
riskier technology and larger costs. Unlike the NF, atmospheric
neutrinos are a significant background since - lowering the energy
-its cross-section weighted rate increases roughly as
$E^{-1.7}$. Suppression of atmospheric $\nu_\mu$ requires the
exploitation of time-correlation between the ion bunch injected in the
decay ring and the time of detection at the far location. In turn,
bunches shorter than $100$~ns are needed, which also represent an
accelerator challenge.  Betabeams producing neutrinos in the
multi-GeV range (Advanced Beta Beams) would open, once more, the
possibility of employing high density detectors and offer LNGS the
opportunity to host a BB far detector. This option (iron calorimeter
in Hall C) has been addressed explicitly in~\cite{LNGS_BB} for a
SuperSPS BB with \Ne\ and \He.  It is less appealing than for the
above-mentioned case of the NF: the somewhat limited mass cannot
compensate for the lower event rate of the BB with respect to the
NF. Moreover, the neutrino energy is still quite small (about 1.5~GeV)
and, therefore, the neutral-current contamination is significant. Larger neutrino
energies (e.g. from \Li8 and \B8) could further enhance its physics
case, although the CERN-LNGS baseline would not be any more at the
oscillation maximum. If the ion yield were appropriate, a natural
solution - of noteworthy strategic interest for INFN and CERN - would
be the use of the SPS with \Li8 and \B8 pointing to an iron
calorimeter located at LNGS. Physics performance are expected to be
comparable with the ones of \cite{LNGS_BB}.

\section{Future scenarios in the standard oscillation framework}

Depending on the value of \thetaot, the experimental approach toward
a full measurement of the $U_{PMNS}$ will be substantially
different.  We discuss below three different ranges of
\thetaot values: very large values, those suggested at 90\% CL by the
global fits of Fogli et al.~\cite{Fit th13} and experimentally
accessible in a couple of years, large values, corresponding to the
ultimate sensitivity of the next generation of dedicated experiments,
say T2K and Daya Bay, accessible in 5-7 years, and small values of
\thetaot, i.e. smaller than the combined sensitivity of the next
experiments.

\begin{figure}
  \centerline{\includegraphics[width=12cm]{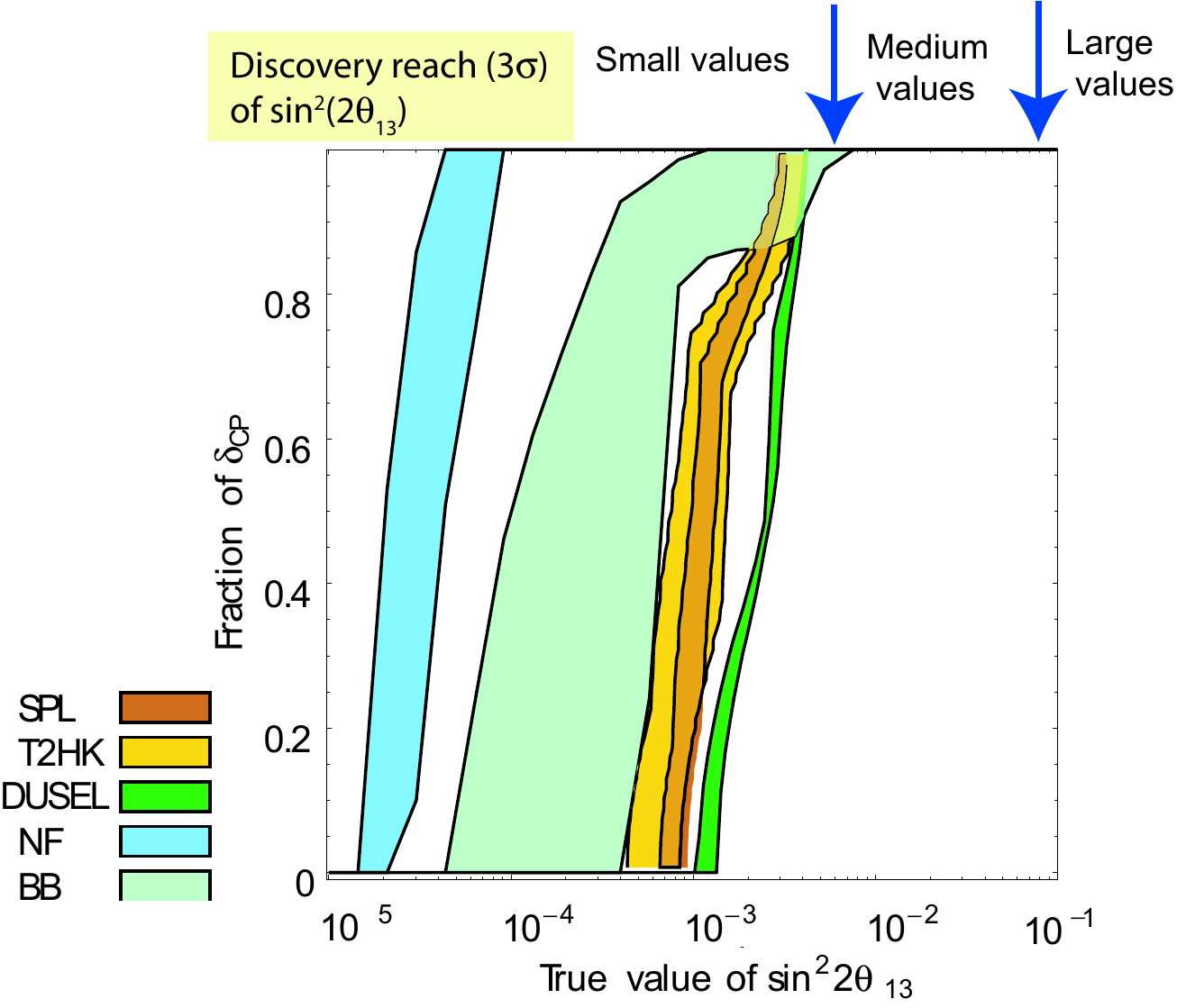}}
  \caption{ The capability to demonstrate that $ \thetaot \neq 0$ at
    3$\sigma$ level for different facilities.  The discovery limits
    are shown as a function of the fraction of all possible values of
    the true value of the CP phase $\delta$ (`Fraction of $\delta_{\rm
    CP}$') and the true value of $\sin^2 2 \theta_{13}$.  The
    right-hand edges of the bands correspond to more conservative
    setups while the left-hand edges correspond to fully optimized
    setups.  The discovery reach of the SPL\index{SPL super beam}
    super~beam from CERN to Frejus is shown as the orange band, that
    of Hyper-Kamiokande (T2HK) as the yellow band, and that of
    Fermilab to DUSEL ``wide band'' beam~\cite{WBB} as the green band.
    The discovery reach of the Beta Beam is shown as the light green
    band and the Neutrino Factory\index{Neutrino factory} discovery
    reach is shown as the blue band.  From~\protect\cite{ISS_phys}. }
  \label{fig:ISS-th13}
\end{figure}

\begin{figure}
  \centerline{\includegraphics[width=12cm]{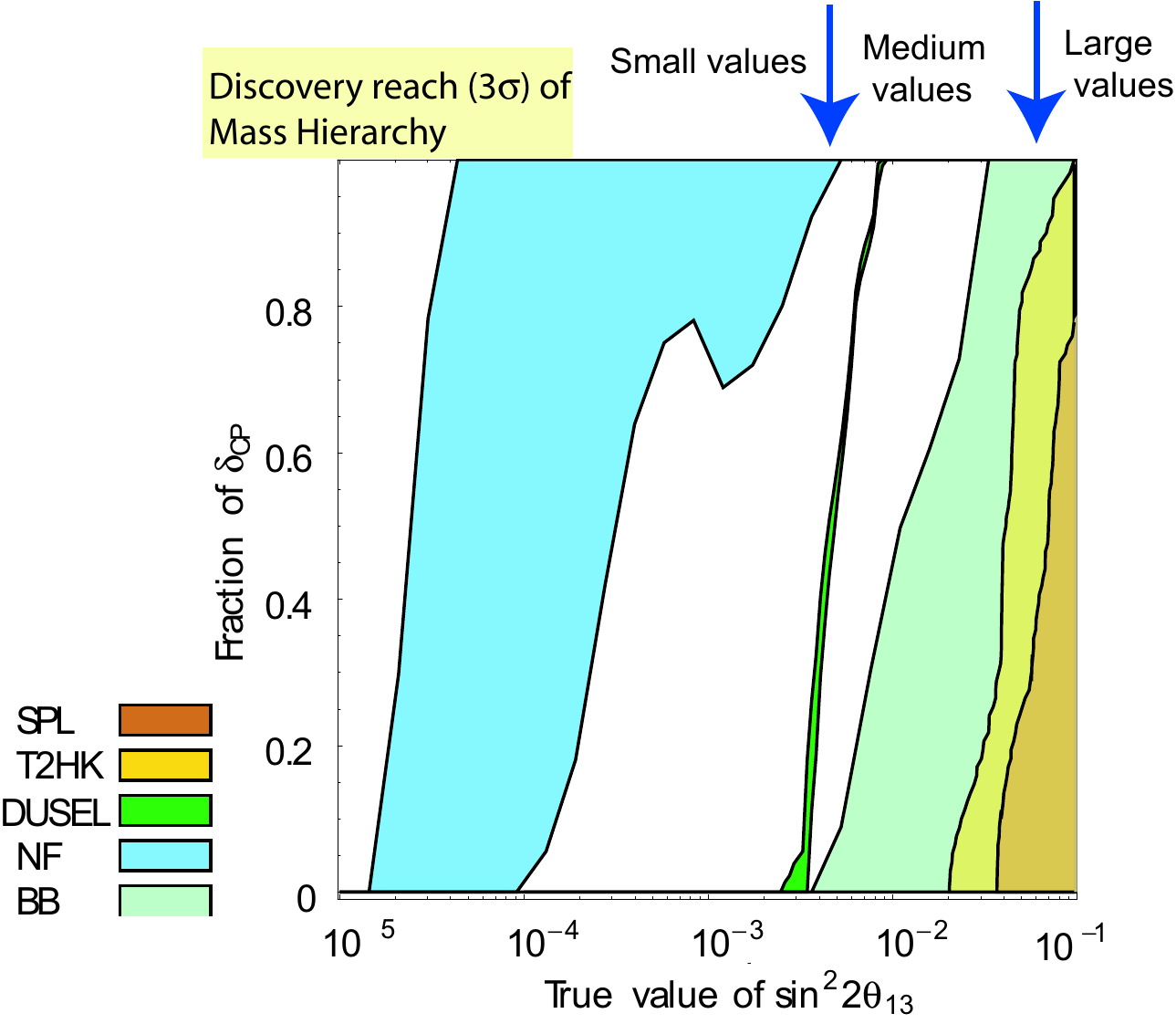}}
  \caption{ The capability to distinguish between normal and inverted
    mass hierarchy at 3 $\sigma$ level for different facilities.  The
    discovery limits are shown as a function of the fraction of all
    possible values of the true value of the CP phase $\delta$
    (`Fraction of $\delta_{\rm CP}$') and the true value of $\sin^2 2
    \theta_{13}$.  The right-hand edges of the bands correspond to the
    conservative setups while the left-hand edges correspond to the
    optimized setups.  The discovery reach of the SPL super~beam is
    shown as the orange band, that of T2HK as the yellow band, and
    that of the wide-band beam experiment as the green band.  The
    discovery reach of the Beta Beam is shown as the light green band
    and the neutrino factory\index{Neutrino factory} discovery reach
    is shown as the blue band.  From~\protect\cite{ISS_phys}.
    }
  \label{fig:ISS-sigdm}
\end{figure}

\begin{figure}
  \centerline{\includegraphics[width=12cm]{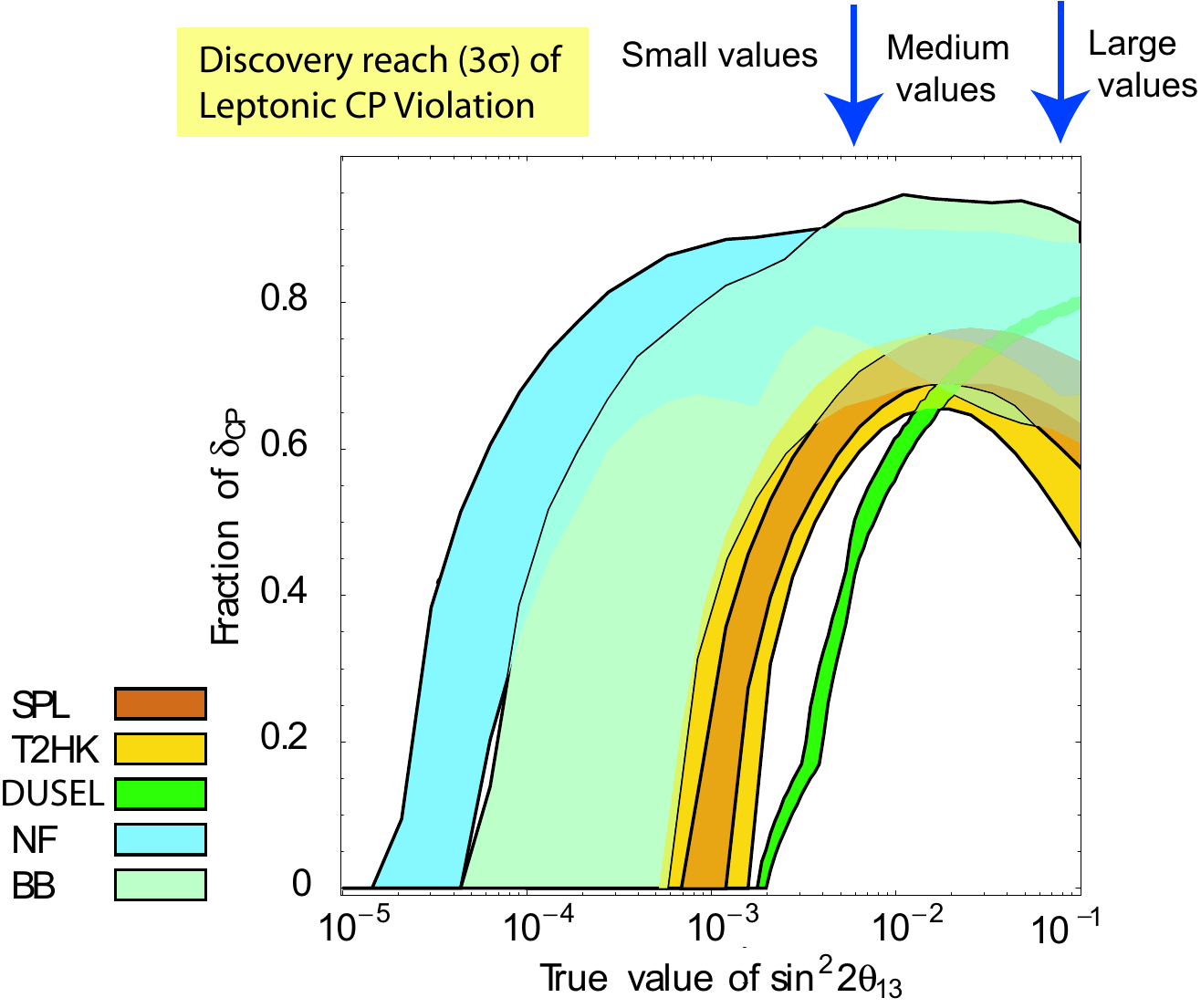}}
  \caption{ The capability to establish CP violation in the leptonic
    sector at 3 $\sigma$ level for different facilities.  The
    discovery limits are shown as a function of the fraction of all
    possible values of the true value of the CP phase $\delta$
    (`Fraction of $\delta_{\rm CP}$') and the true value of $\sin^2 2
    \theta_{13}$.  The right-hand edges of the bands correspond to the
    conservative setups while the left-hand edges correspond to the
    optimized setups.  The discovery reach of the SPL super~beam is
    shown as the orange band, that of T2HK as the yellow band, and
    that of the wide-band beam experiment as the green band.  The
    discovery reach of the Beta Beam is shown as the light green band
    and the neutrino factory\index{Neutrino factory} discovery reach
    is shown as the blue band.  From~\protect\cite{ISS_phys}.
    }
  \label{fig:ISS-LCPV}
\end{figure}

To discriminate among the options, we use as a reference the
sensitivity plots computed within the aforementioned (see
Sec.~\ref{sec:NF}) International Scoping Study (ISS) \cite{ISS_phys}.
Figs. \ref{fig:ISS-th13},\ref{fig:ISS-sigdm},\ref{fig:ISS-LCPV}
display the expected sensitivity of the facilities discussed in
Sec.~\ref{sec:new_facility} - Super Beams, Beta Beams and Neutrino
Factories - in measuring \thetaot, \sigdm \, and $\delta$ as a function
of \thetaot.  Since the performance of the facilities have large
uncertainties, they are displayed as wide areas under different
assumptions on the final layout and on the limiting
systematics\footnote{For details, we refer the reader to the first
volume of the full ISS report \cite{ISS_phys}. A comprehensive
description of the detector options and accelerator performance are
available in Vol. II \cite{ISS_det} and Vol. III \cite{ISS_acc},
respectively.}. Note that in some cases, the width of
the area can be even larger than one order of magnitude. It is worth
stressing that so far no realistic estimate of costs and timescales of
Neutrino Factories and Beta Beams exists: it is one of the prime tasks
of the European Network EURO$\nu$ (see Footnote 7) to reach firm
conclusions on this item.

\subsection{Very large \thetaot ($\thetaot \geq 8^\circ$ or $\sin^2{2\thetaot} \geq 0.08$)}

If \thetaot is just below the CHOOZ limit, it will be measured in the
first runs of T2K and Double-Chooz, hopefully within 2010.  For so
large \thetaot , first results on leptonic CP violation might be
reached by upgrading the beam intensity of the existing
facilities (T2K and NO$\nu$A) with no new gigantic far detectors
\cite{Huber:2009cw}.  It is easily predictable that such upgrades
will have a very high priority in Japan and USA and in the meantime a
megaton water Cerenkov detector (or 100 kton liquid Argon detectors)
could likely be funded at Kamioka and/or DUSEL.

In such a scenario a major investment in Europe for a new
infrastructure would be fully justified. In Italy, it might bring to
the construction of a new shallow-depth laboratory to host LAr TPC's
of a size not accessible at LNGS.  This option has been discussed in
the framework of the ModulAr proposal~\cite{modular}, where a 20~kton
detector would be installed slightly off the main axis of CNGS, in a
way that resembles the location of NOVA along the
Fermilab-to-Minnesota neutrino beam (see Fig.\ref{fig:nova}). 
In general, the use of the imaging capability of the LAr-TPC ensures a
higher discovery potential than in the case of scintillator (or water)
detectors, i.e. a comparable sensitivity may be achieved with a
smaller sensitive mass.

From the point of view of LNGS, large values of
\thetaot will bring measurable signals in OPERA.  The OPERA
result would be very interesting: it is truly complementary to the T2K
and NOVA results since the facility is operated off the peak of the
oscillation maximum and with completely different
systematics~\cite{opera_numunue,migliozzi}.  As a consequence it would
be very recommendable a prolongation of the OPERA run, possibly with
an intensity upgrade of the CNGS.

\subsection{Medium values of \thetaot ($\thetaot \geq 2^\circ$ or $\sin^2{2\thetaot} \geq 0.005$)}
In this range of values a non-zero \thetaot value will be established
at 3$\sigma$ in the full run of T2K and Daya Bay (around 2015).  For
these values no simple upgrades of T2K and NOVA will have sensitivity
on CP violation and mass hierarchy (we include ModulAr in this class
of experiments), and detectors of the class of Hyper-Kamiokande will
be needed.  This is a favorable scenario for the Superbeams: they
represent a relatively low risk technology and can get a clearly
accessible physics case for medium values of \thetaot.  In Europe the
SPL project, with a megaton class detector installed in a new cavern
at the Frejus Laboratories, could compete with these projects, having
a better sensitivity on \thetaot and CP violation and a worse
sensitivity on mass hierarchy \cite{memphis}. In this race, Europe
suffers from some disadvantages. It is likely that only one
megaton-size detector will be funded world-wide. Japan is a leader on
this technology and US would likely exploit the discovery of \thetaot
to boost this opportunity at  DUSEL. Moreover, at that time (2016) CERN
will be engaged with the replacement of the PS and, probably, with the
luminosity upgrades of the LHC, so funding for an ultimate facility
for astroparticle physics might be difficult to be gathered to
successfully face a competition with US and Japan.

As noted above, however, Europe could pursue better suited
strategies than the standard Superbeam.  If the feasibility of a
high-Q SPS-based Betabeam can be demonstrated in a few years from now,
CERN could potentially host a novel facility for neutrino oscillations
exploiting either smaller detectors (high density calorimeters in Hall
C) or with significantly better performance if combined with a
100~kton size LAr TPC.  According to \cite{Winter:2008dj}, in this
range of values the optimal facility (the one that could measure
\thetaot, \sigdm and $\delta$ with just one detector) would be a
high-$\gamma$ (or even an SPS based) Beta Beam pointing to a megaton water
Cerenkov detector or a detector with equivalent performance (e.g. a
100 kton LAr TPC).

There is however a more conservative approach that might be extremely
interesting for UK and Italy. The guarantee of the physics case due to
the finite size of \thetaot can be used to strongly boost the
realization of a Neutrino Factory based at the Rutherford Labs. As a consequence, CP
violation and matter effect could be studied in the experimental halls of LNGS building a
iron magnetized detector (see Sec.~\ref{sec:NF}) located about 1500~km
from the source.  Ultimate precision for \thetaot and for the resolution of
the degeneracies will be achieved exploiting the second RAL-to-India
baseline (7000~km).  

\subsection{Small values of \thetaot ($\thetaot \leq 3^\circ$ or $\sin^2{2\thetaot} \leq 0.01$, i.e. 
null results from the next generation of experiments)}

If no signal of \thetaot will be detected by the next generation of
 accelerator and reactor experiments, a rather problematic scenario
 will disclose.

The discovery potential of Super Beam experiments like Hyper-Kamiokande
and DUSEL would be quite limited (they have $3\sigma$ CP violation
sensitivity for \thetaot values as small as about $\sin^2{2\thetaot}
\leq 0.005$) and it seems very unlikely that those facilities will be
built under these circumstances.  Advanced Beta Beams and Neutrino
Factories can explore \thetaot values two or three orders of magnitude
smaller than the T2K sensitivity, but again it seems unlikely that
facilities of such cost will be strongly supported having as a physics case
just the exploration of the \thetaot value and a search for CP
violation without any guaranteed signal.  On the other hand it seems
unlikely that neutrino oscillation experiments stop without
mapping the values of three fundamental parameters of the Standard
Model: \thetaot, \sigdm \, and $\delta$, also considering that in the
framework of Leptogenesis $\delta$ is the single most important
unknown ingredient to address matter-antimatter asymmetry in the
Universe \cite{Leptogenesis}.
In this case, however, any progress in oscillation neutrino physics will rely on
advances in Beta Beams, Neutrino Factory or further novel sources.


\section{Neutrino oscillation physics beyond the standard framework}
\label{sec:nonstandard}

Among elementary fermions, neutrinos have a unique status due to
charge neutrality and the extremely small value of their
mass. Most likely, they are the only Majorana particles up to the
electroweak scale and, differently from other oscillating particles,
their coherence length can extend up to cosmic
scales~\cite{farzan}. It comes to no surprise that these particle have
been speculated to have unusual properties.  Within several models,
their free streaming in vacuum and matter is perturbed by non-standard
interactions, which can be constrained mainly by oscillation
experiments. Even Lorenz-invariance and CPT conservation~\cite{CPT}
have been challenged for neutrinos.  As a matter of fact none of these
hypotheses are empirically grounded and the constraints coming from
the oscillation experiments and from cosmology are extremely
tight. However, a further stimulus to pursue these searches has
come from the persistent LSND anomaly, which mainly brought to the
consideration of additional sterile neutrinos.

LSND indicated \cite{lsnd} a $\bar \nu_\mu \to \bar
\nu_e$ oscillation with a third, very distinct, neutrino mass
difference: $\Delta m_{LSND}^2 \sim 0.3 - 6$ eV$^2$
\footnote{The LSND experiment indeed wasn't able to define the range of
$\Delta m^2$; what is quoted here is the region not completely excluded by other
experiments.}. Already in 1999,
Fogli et al. \cite{Fogli:1999zq} showed that the standard three
neutrino framework is inconsistent with the ensemble of all
experimental data, included LSND.  To explain the whole set of data at
least four different light neutrino species would have been needed
(referred as ``3+1'' models).

 Sterile neutrinos are
particles with the same quantum number of the vacuum and, therefore,
are ``sterile'' with respect to electroweak interactions. In the
standard three-family framework, massive sterile neutrinos are the ordinary
right-handed components of the Dirac fermions and they do not mix with
active neutrinos. Beyond this model, light steriles could mix with
active $\nu$ and, therefore, change the transition probability through
a $N \times N$ leptonic mixing matrix with $N>3$~\cite{Volkas}.  
These model predicted of course several other oscillation processes in
excess to the $\nubarmu-\nubare$ transitions detected by LSND, the largest
part already excluded by other short baseline experiments. As a consequence,
these $3+1$ models provided anyway poor overall fits to the existing
experimental data \cite{Maltoni:2004ei}.

 The MiniBooNE experiment at FermiLab
was designed to directly check the LSND claim, performing a search for
$\nu_\mu \to \nu_e$ appearance with a baseline of 540 m and a mean
neutrino energy of about 700 MeV, i.e. with energy and baseline larger
than LSND but with the same ratio $L/E$. In 2007 MiniBooNE
published a no-evidence analysis for \numunue oscillation \cite{miniboone2007}.
While this analysis was not able to rule-out the whole parameter space
of the LSND signal~\cite{Karagiorgi:2009nb}, it represented a further
suppression of the four-neutrino interpretation of the LSND
anomaly.
 In 2009 MiniBooNE~\cite{miniboone2009} published also results for
\nubarmunubare \, oscillation showing, again, no appearance effects,
although at a smaller significance with respect to the neutrino data.

MiniBooNE neutrino data reported also an unexplained excess for low energy
neutrinos (MineBooNE anomaly in the following), that cannot be
reconciled with any sterile model. This data excess was not present in the
antineutrino run.

At present, electron (anti)neutrino appearance results, including
the LSND and the MiniBooNE anomalies, can be reconciled in a ``3+2'' model
\cite{Maltoni:2007zf}. However results  from muon neutrino disappearence
experiments at short baselines are in disagreement with
this model \cite{Karagiorgi:2009nb,Maltoni:2007zf} so that the
 world dataset are clearly inconsistent with both 3+1 and 3+2 model
predictions and even more baroque models should be advocated. 

It is interesting to analyze whether the CNGS might be able to test
the LSND anomaly, in spite of the much larger $L/E$. Naively, if the LSND
anomaly comes from a two family oscillation source of $\nu_\mu
\rightarrow \nu_e$ parametrized as $\sin^2 2\theta_{LSND} \cdot \sin^2
(\Delta m^2_{LSND} L/4E)$ with $\Delta m^2_{LSND} \simeq 1$~eV$^2$~\footnote{This
formula would be plausible for instance if you did not need two
different mass scales to explain the solar and atmospheric neutrinos:
in this case $\Delta m^2_{LSND}$ could play the role of \dmot\ or
\dmtt. Although oversimplified, this formula is still used to express
the sensitivity coverage of KARMEN~\cite{karmen} and MiniBooNE with respect to the
LSND claim.}, at CNGS the probability of $\nu_\mu \rightarrow \nu_e$
oscillation would be very high (i.e. $0.5\sin^2 2\theta_{LSND}$) since the
oscillating term averages out to 1/2. Moreover, thanks to the increase
of the cross section at 17~GeV, the CNGS would perform significantly better than
MINOS or other long-baseline experiments.  Unfortunately, if any
realistic model (e.g. 3+1) is used to compute the actual rate at LNGS,
the effect is rather small, since models must evade the already
stringent bounds from atmospheric neutrinos.  A detailed calculation of the
performance of the CNGS to test the LSND anomaly has been done
in~\cite{sterili}. In general, we can expect that the CNGS will
improve marginally the bounds on the sterile parameters and,
surprisingly, much more from the \numunutau\ analysis than from the
\numunue.  On the other hand, \numunutau\ oscillations are particularly
suited to test unconstrained parameters for the non-standard interactions. It has been
shown, for instance, that operators affecting the \numunutau\
transitions and compatible with current constraints from atmospheric
might change substantially the number of observed tau events in
OPERA~\cite{NSI}. It is, however, in the framework of the neutrino
factories that a superior test of non-standard interactions (at 1-0.1 \% level) can be
carried out either with or without a detector for the silver
channel~\cite{NSI_winter}.

Given the present experimental situation, the most straightforward way
to test (again) LSND is to perform a similar experiment at the same L
and E but with significantly larger statistics. An experiment based on
large fluxes of pions decaying at rest and in flight has been recently
proposed at the new US neutron spallation source (SNS)~\cite{oscSNS}.

Another straightforward way would be to incorporate a close detector
in the MiniBooNE setup, greatly reducing the impact of systematic errors
in the experimental analysis.
The recent BooNE letter of intents~\cite{Boone} showed that
building a second MiniBooNE detector at (or
moving the existing MiniBooNE detector to) a distance of $\sim 200$
 m from the Booster Neutrino Beam (BNB) production target, the
sensitivity of the setup would result greatly enhanced.

Similarly, the MiniBooNE low-energy anomaly - which is located at a
different E and L/E with respect to LSND - might be studied directly
using a high granularity detector positioned along the same beam of
MiniBooNE (BooNE beamline) at Fermilab
(``MicroBooNE''~\cite{microboone}). This detector, operated both at
BooNE and at NuMI could simultaneously test the above-mentioned
anomaly and perform high precision cross section measurement in the
critical region around 1~GeV. This region is of great practical
interest since most of the Superbeams proposed so far are operated in
such energy range; it is explored by a dedicated cross-section
experiment called SciBooNE~\cite{sciboone} and, along NuMI, by
Minerva~\cite{minerva}.  The LAr technology is ideal to perform these
measurements and, therefore, it has been proposed as the basic
technology option for MicroBooNE up to mass of 170~t (70~t fiducial
mass).

 As discussed in \cite{rubbia_cern,doublelar}, an even larger
detector of 500~t active mass could be built at CERN restoring the
low-energy neutrino beamline driven by the PS and originally used for
BEBC-PS180~\cite{BEBC}. The detector would be positioned 850~m far
from the source and complemented by a smaller near detector at 127~m
(150~t active mass).  An identical setup, with the notable difference
of a totally active scintillator detector instead of liquid
argon, was proposed to CERN already in 1997 \cite{I216}.  With respect
to MicroBooNE, the presence of an identical near detector strengthen
significantly the reliability of the oscillation search and the
overall mass allows for a test based on a much larger statistics. On
the other hand, this setup requires the construction of a dedicated
beamline and proper sharing of the protons of the PS, which also feed
the rest of the CERN physics programme (SPS fixed target and the
LHC). The CERN option (1.1 GeV at 850~m) works, once more, at the same
L/E as LSND. The values of L and E, however, are different both with
respect to the original LSND apparatus and also with respect to
MiniBooNE (0.7~GeV at 550~m). This situation is ideal if the source of
anomaly has the usual L/E pattern but could be non optimal if the LSND
and/or the MiniBooNE excess has a different origin.

\section{Conclusions}

In this paper, we investigated the perspectives for accelerator
neutrino physics in Europe, with special emphasis to the line of
research where INFN infrastructures (firstly LNGS) and expertise could
be most profitably exploited. If all further data coming from the
current and next generation of experiments fit the standard oscillation
framework, i.e. the assumption of three active neutrinos with Standard Model
couplings, then the strategy for the next decade will be solely driven
by the actual size of \thetaot. In particular:

\begin{itemize}
\item if \thetaot is close to the current limits ($>8^\circ$) and
therefore can be observed very early by T2K or Double-Chooz, an
independent measurement done in parallel by the CNGS experiments would
be of high value: it could be obtained by high granularity detectors
with completely different systematics with respect to T2K and,
moreover, running off the peak of the oscillation probability, with no
dependence on matter effect and with a dependence of the cosine of the
CP phase instead of the usual $\sin \delta$ term. Therefore, a
prolongation of the CNGS, possibly running in dedicated mode or
profiting from an intensity upgrade should definitely be pursued.
Moreover, in this special case medium size detectors (T2K, NOVA) with
upgraded beams could play a relevant role for precision measurement
in the leptonic sector. A medium size (20~kton) LAr detector - having
performance comparable or better than NOVA - could complement the T2K
measurements, especially with the aim of determining the mass
hierarchy of neutrinos through the observation of matter
effects. Clearly, it is an ideal scenario for CERN and INFN, who would
fully profit of the investments done for CNGS even beyond the primary
aim of the experiment, i.e. the observation of \numunutau\
oscillations.

\item for lower values of \thetaot, several options can be envisaged
both inside and outside the Gran Sasso Laboratories.  In a timescale
shorter than the Neutrino Factory, a SPS-based Betabeam at CERN accelerating
high-Q ions might become available. In this case, an experimental hall of the
Gran Sasso labs would be enough to host a high density detector
(e.g. iron calorimeter) operating in \numu\ appearance mode. A high-Q
Betabeam requires a significant R\&D which, however, is synergical
with the activities of nuclear physics community and it surely
exploits in a clever manner European infrastructures, as CERN, the
INFN Laboratories of Legnaro, CRC at Louvain and several other labs working
on the development of radioactive ion beams.

On the other hand, in this range of \thetaot, the construction of the
Neutrino Factory is considered as the ultimate and most precise
facility for the study of lepton mixing. LNGS could host the far
detector of a UK-based Neutrino Factory although the distance from the
Rutherford Labs to LNGS is shorter than the optimal one (1500 versus
3000~km).  This option offers the best physics performance at a price
of a longer timescale since it is unrealistic that a Neutrino Factory
comes into operation before 2020.

In the context of a CERN-based or RAL-based Neutrino Factory
($>2020$), an option highly synergical with INFN infrastructures and
technologies is the study of $\nue \rightarrow \nutau$ transitions at
LNGS using detectors able to identify the appearance of tau leptons,
as it was the case of the CNGS.  This ``silver channel'' can be
exploited at LNGS by a $\sim$5~kton OPERA-like or ICARUS-like detector
equipped by a magnetic spectrometer for $\mu$ charge measurement and
detecting neutrinos from CERN or RAL.  This channel can be used to solve
the degeneracies in the mixing parameters when measured with an
ultimate facility as the Neutrino Factory itself.

Without Betabeams or Neutrino Factories, there is no way to use the
halls of LNGS profitably over this range of \thetaot.
As an alternative (or in parallel, if resources and interest from
other countries can be gathered), the European neutrino community
could pursue an aggressive R\&D to boost the LAr technology to the
100~kton size: an ultimate facility for astroparticle physics and for
the technology of the Superbeam.

\item in case of no evidence of \nue appearance after T2K, it is
unlikely that accelerator neutrino physics will gather major
investments. R\&D toward the Neutrino Factory will proceed at a
lower rate and will be done synergically with a possible Muon Collider
while a megaton-size Water Cherenkov detector or a 100~kton LAr TPC
will be mainly motivated by astroparticle physics and proton decay
searches.
\end{itemize}

\noindent
Outside the standard framework, we still see  some opportunities
especially for ``low'' mass Liquid Argon detectors in the quest for the
clarification of the low energy excess of MiniBooNE. The most natural
application is in the framework of MicroBooNE~\cite{microboone}: an
experiment based on a $\sim100$~ton LAr TPC and located at Fermilab,
along the BooNE and NuMI beamline.  Alternatively, as pointed out in
\cite{rubbia_cern}, the PS neutrino beamline at CERN could be restored to
perform a similar test with better significance, and with the added
value of the continuity in the development and exploitation of the
LAr technology within Europe. Note, however, that in a few year
a further test~\cite{oscSNS} of the original LSND anomaly might also be
done at the SNS, the US high intensity neutron source already in
operation at the Oak Ridge National Labs.

As a final consideration, we wish to stress that these years - which
have witnessed the completion and the startup of the CNGS and, more
recently, of Double-Chooz - are unique times to look to the future of
neutrino physics with artificial sources in
Europe~\cite{workshopCERN}.  Clearly, the optimal strategy will be
finally driven by the outcomes of the \thetaot-oriented experiments,
but a systematic exploration of the different alternatives as a
function of these outcomes was missing in literature, especially when
focusing to the existing European infrastructures and to technologies
where INFN has a world leadership: we hope that this paper fills such
gap at best of our present knowledge of lepton mixing.

\section*{Acknowledgments}
This work originated within the ``Commissione Scientifica Nazionale
II'' (Astroparticle Physics) of INFN and EURO$\nu$.  We thank our
colleagues of Commissione II and the Director of the LNGS Labs (Lucia
Votano) for many enlightening discussions during the drafting of the
paper.  We also want to express our gratitude to Carlo Rubbia who
provided us detailed information on the Double-LAr
project~\cite{doublelar} at CERN-PS.  Finally, we gratefully
acknowledge D.~Gibin, A. Guglielmi, F. Pietropaolo, C.~Rubbia and
P. Sala for critical reading of the manuscript and many interesting
suggestions.

  \end{document}